\documentclass[12pt]{article}

\usepackage[dvips]{graphicx}
\usepackage{amsmath}
\usepackage{amsfonts}
\usepackage{amssymb}
\usepackage{hyperref}
\usepackage{color}

\newcommand{\be}{\begin{equation}}
\newcommand{\ee}{\end{equation}}
\newcommand{\bea}{\begin{eqnarray}}
\newcommand{\eea}{\end{eqnarray}}

\newcommand{\nn}{\nonumber}

\newcommand{\lp}{\left(}
\newcommand{\rp}{\right)}

\begin{document}

\begin{flushright}
\small
DESY 13-060 \\
CERN-PH-TH/2013-062\\
\end{flushright}

\vskip 8pt

\begin{center}
{\bf \LARGE {
 On the non-linear scale of\\
 \vskip .4cm
  cosmological perturbation theory
 }}
\end{center}

\vskip 12pt

\begin{center}
\small
{\bf Diego Blas$^{a}$, Mathias Garny$^{b}$, Thomas Konstandin$^{b}$ } \\
\vskip 3pt
{\em $^a$ CERN, Theory Division, 1211 Geneva, Switzerland } \\
{\em $^b$ DESY, Notkestr.~85, 22607 Hamburg, Germany } \\
\end{center}

\vskip 20pt

\begin{abstract}
\vskip 3pt
\noindent
 We discuss the 
  convergence of cosmological perturbation theory. We prove that 
 the polynomial enhancement of the non-linear corrections expected from the effects
 of soft modes is
 absent in equal-time correlators like the power or bispectrum.
We first show this at leading order by resumming
the most important corrections of soft modes to an arbitrary skeleton of hard
fluctuations. We derive the same result in the eikonal
approximation, which also allows us to  show the absence of enhancement
 at any order. We complement the proof by an explicit calculation of the power spectrum at two-loop order,
and by further numerical checks at higher orders. Using these insights, we 
argue that the modification of the power spectrum  from soft modes 
corresponds at most to logarithmic corrections at any order in perturbation theory. Finally, we 
discuss the asymptotic behavior
in the large and small momentum regimes and identify the expansion parameter pertinent
to  non-linear corrections.
\end{abstract}

\newpage

\section{Introduction\label{sec:intro}}

The theory of cosmological perturbations is the standard tool to understand the emergence of 
the large scale structure of the Universe \cite{Peebles:Book,Bernardeau:2001qr}. This approach
is based on the assumption that small perturbations around an otherwise homogeneous and isotropic
Universe grow with time due to their gravitational interaction. This growth is particularly efficient
for scales inside the Hubble horizon and in the matter-dominated epoch \cite{Peebles:Book}. 
The tiny amplitude of the primordial perturbations allows for a perturbative treatment
of this non-linear problem, but the aforementioned growth eventually invalidates
this approach. The scale at which this happens has been traditionally identified with
moments of the linear power spectrum. 
A particularly interesting scale is created by the enhancement 
of non-linearity  coming from the 
coupling of modes of small momenta (\emph{soft}) to the modes of the
scale of interest (\emph{hard}), 
\be
\label{eq:kNL}
k_{NL}^{-2}\sim \int d q P^L(q,\eta)\,,
\ee
where $P^L(q,\eta)$ is the linear power spectrum at time $\eta$ (we will be more precise about this point below).
To study physics beyond this scale, it seems unavoidable to resort to non-perturbative schemes. 
Particularly well-suited to deal with this effect is the 
 scheme known as renormalized cosmological perturbation theory (RPT) \cite{Crocce:2005xy}. 
This  systematic approach to
cosmological perturbation theory is based on the introduction of the non-linear propagator, for which 
the non-linear effects associated to  the scale (\ref{eq:kNL})  were derived in \cite{Crocce:2005xz}.
These result in an exponential suppression controlled by $k_{NL}$ that was also conjectured
to be the leading effect in the matter power spectrum. 

The seminal work  \cite{Crocce:2005xy} was followed by 
many other resummation schemes  put forward to cope with non-linear effects 
 in an analytical way (i.e. without
resorting to large $N$-body simulations)
 \cite{Taruya:2007xy,Matarrese:2007wc,Valageas2007N,Matsubara:2007wj,Pietroni:2008jx,Anselmi:2012cn}.
The hope is that after resummation of a subset of diagrams the resulting expansion is under better control.
For schemes related to the effects of the scale (\ref{eq:kNL}) in the power spectrum 
this expectation is at odds with two results. 
First, it is known since a long time that the leading contributions to the power spectrum from soft modes at arbitrary loop order cancel~\cite{Jain:1995kx,Scoccimarro:1995if}. In the above mentioned resummation schemes this cancellation is normally not explicit. 
Second, one can try to systematically 
understand the effects of soft modes on hard modes  by using the \emph{eikonal} approximation
\cite{Bernardeau:2011vy}. In this case, it was shown that the mentioned suppression for the propagator is present
while the power spectrum is unchanged \cite{Bernardeau:2012aq}. 
A first aim of this work is to reconcile both approaches and understand how the eikonal result can be recovered  from the 
diagrammatic technique of resummation of soft modes.

 The cancellation 
of the effects from soft modes suggests that the convergence properties of standard perturbation theory (SPT) 
are not ameliorated by resuming the soft modes. It also suggests that the effects in the power spectrum associated 
to $k_{NL}$ are  spurious,  
which immediately rises the question of what is the real parameter governing 
the non-linear corrections for this observable, also in SPT. It is simpler to answer this point after
 implementing the eikonal approximation in a controllable way. We will do this in the second part of the paper. 
The scale (or parameter) controlling the non-linear dynamics in this case, and not (\ref{eq:kNL}), should be the one
tamed by any resummation scheme with a better validity than
the standard case for the non-linear dynamics.

Our work is organized as follows:
Section~\ref{sec:notation} clarifies our notation and reviews standard perturbation theory (SPT). Section~\ref{RPT_eik} discusses the two main resummation schemes we are concerned with, namely renormalized perturbation theory (RPT) and the eikonal approximation. In section~\ref{skeletons_resummed}
we generalize the resummation of RPT to a larger class of soft corrections that are relevant to the power spectrum and other equal-time correlators. This motivates the discussion of next-to-leading order corrections in the eikonal approximation presented in section~\ref{sec:eikonal}. The main result is that no enhancement from soft vertices should be present in equal-time correlators. We further support these claims by explicit analytic and numerical results presented in section~\ref{power_spec}. We conclude in section~\ref{sec:discussion}. Some technical details are relegated to the Appendices.

\section{Standard cosmological perturbation theory  
  \label{sec:notation}}

In this section we set up the system of equations relevant for our discussion. 
As stated in the introduction, we are interested in understanding  
some features about the behavior of cosmological perturbations
when their amplitude grows to the point that non-linear corrections
are important. 
For this problem it is enough to consider sub-horizon perturbations in a matter-dominated
era. We will also assume\footnote{Deviations from this 
assumption can be taken into account by using the effective language
of \cite{Carrasco:2012cv} (see also \cite{Pietroni:2011iz}). Since we focus on the first non-linear
effects those deviations are not important for our results.} that the matter in the Universe
is a perfect fluid described by a density field $\rho(x,\tau)=\rho(t)(1+\delta(x,\tau))$
and a velocity field $v^i(x,\tau)$
defined at a conformal time $\tau\equiv \int d t /a(t)$.
In this case the Newtonian cosmological perturbations
yield an accurate  description \cite{Peebles:Book}. Finally, 
we will reduce our analysis to the case without vorticity where $\theta\equiv \partial_i v^i$.
In this case, the
perturbations can be written in a two-component form \cite{Bernardeau:2001qr,Scoccimarro:1997gr}, %
\be
\label{eq:Psi_eta}
\partial_\eta \Psi_a(k,\eta) + \Omega_{ab} \Psi_b (k,\eta) = 
\gamma_{abc}(k_1,k_2) \Psi_b(k_1, \eta) \Psi_c(k_2, \eta) \, ,
\ee
where 
\be
\Psi_1(k,\eta)\equiv \delta(k,\eta), \quad \Psi_2(k,\eta)\equiv -\frac{\theta(k,\eta)}{f_+(\eta)\cal{H}},
\ee
and we have introduced the functions ${\cal H}\equiv \frac{d\ln a(\tau)}{d \tau}$ 
and $f_+(\eta)\equiv \frac{d \ln D_+(\eta)}{d \ln a(\eta)}$
and the  time $\eta\equiv \ln D_+(\tau)$, with $D_+(\tau)$ being the 
linear growing mode of the density contrast
\be
\delta^L(k,\tau)=D_+(\tau)\delta_0(k).
\ee
The momenta $k, k_1, k_2$ are vectors and the equation is understood as being
summed over double indices and integrated over the two momenta $k_1$ and $k_2$
with the measure $\delta^{(3)}(k-k_1-k_2)$ on the right-hand side. 
This  fixes our convention for Fourier transforms.
In the following we
only write indices and integrations when the notation of an equation could be
ambiguous.

The matrix $\gamma$ constituting the mode coupling can be written in symmetric
form with the elements
\bea
\gamma_{121} = \alpha(k_1, k_2)/2, \quad
\gamma_{112} = \alpha(k_2, k_1)/2, \quad
\gamma_{222} = \beta(k_1, k_2), \,
\eea
with
\bea
\alpha(k_1, k_2) &\equiv& \frac{(k_1 + k_2) \cdot k_1}{k_1^2}, \quad \quad
\beta(k_1, k_2) \equiv \frac{(k_1 + k_2)^2 k_1 \cdot k_2}{2 k_1^2 k_2^2} \, ,
\eea
and all other elements vanishing. 

In the case where one of the momenta flowing into the vertex is soft ($k_1 \ll k$ or $k_2 \ll k$) 
the previous vertex reduces to 
\be
\label{eq:soft_vertex}
\gamma_{ijk} \to  \delta_{j2}\delta_{ik} \frac{k_2 \cdot
k_1}{2 k_1^2} + 
\delta_{k2}\delta_{ij} \frac{k_2 \cdot k_1}{2 k_2^2} \, .
\ee
The matrix $\Omega$ depends on the underlying cosmology and can depend on
the conformal time $\eta$ (but not on momentum). It may also contain information
about modifications of gravity \cite{Pietroni:2008jx}.
Remarkably, our main results will be valid for any $\Omega$. For completeness
let us remind that
in a flat matter-dominated Einstein-de Sitter Universe for which $\Omega_m=1$
and $D_+(\tau)=a(\tau)$
it reads
\be
\label{eq:EdSOm}
\Omega = \begin{pmatrix}
0 & -1 \\
-3/2 & 1/2 \\
\end{pmatrix} \, .
\ee
Note also that the
 common  Zel'dovich approximation \cite{Zeldovich:1969sb} differs from the exact dynamics
(\ref{eq:Psi_eta}) only by
 the choice of the $\Omega$-matrix~\cite{Bernardeau:2001qr,Valageas:2007ge,Hui:1995bw}
\be
\label{eq:OZA}
\Omega^{ZA} = \begin{pmatrix}
0 & -1 \\
0 & -1 \\
\end{pmatrix} \, .
\ee
The equations (\ref{eq:Psi_eta}) can be solved perturbatively by treating the right side of the
equation that mixes modes with different momentum as a perturbation to the linear equation
\be
\partial_\eta \Psi^L(k,\eta) + \Omega(\eta) \Psi^L(k,\eta) = 0 \,\label{eq:linear} .
\ee
This perturbative scheme, known as standard perturbation theory or SPT, is based on the assumption 
that the amplitude of the perturbations $\Psi$ is small. The solution to  the previous equation can be easily written in term of the Green's function and the initial conditions $\Psi(k,\eta_0)$   as
\be
\label{eq:PsiL}
\Psi^L (k,\eta) =  e^{-\int_{\eta_0}^\eta d\tilde\eta \, \Omega(\tilde\eta)} \,
\Psi(k, \eta_0) \Theta(\eta-\eta_0)\equiv g(\eta, \eta_0) \Psi(k,\eta_0) \, ,
\ee
where we used the notation $\Theta(x)$ for the Heaviside step function to avoid
confusion with the velocity field $\theta$.
If the matrix $\Omega$ is independent of  $\eta$,
the Green's function depends only on the difference $\eta-\eta_0$. For the case (\ref{eq:EdSOm}), its explicit form is 
\be
\label{eq:Green}
g(\eta, \eta_0) = \frac{e^{(\eta - \eta_0)}}{5} 
\begin{pmatrix}
3 & 2 \\ 
3 & 2 \\
\end{pmatrix}\Theta(\eta-\eta_0)
+ \frac{e^{-3(\eta - \eta_0)/2}}{5} 
\begin{pmatrix}
2 & -2 \\ 
-3 & 3 \\
\end{pmatrix}\Theta(\eta-\eta_0) \, ,
\ee
from where we can readily identify the growing and the decaying mode. 
The solution to the equation (\ref{eq:Psi_eta}) can
formally be written as
\bea
\Psi_a(k,\eta) &=& g_{ab}(\eta, \eta_0) \Psi_b(k,\eta_0) \nn \\
&& \quad + \int_{\eta_0}^{\eta} g_{ab}(\eta ,
\bar \eta) 
\gamma_{bcd}(k_1, k_2) \Psi_c(k_1, \bar \eta) \Psi_d(k_2, \bar \eta) \, .
\label{eq:Psi_int}
\eea
The perturbative solution (in powers of $\Psi_a(k,\eta_0)$) can then be obtained by successively reinserting the
left-hand side into the  previous integral. 
Physical observables can be evaluated by performing
statistical averages over these fields. A particularly useful one is the
power spectrum given by 
\be
\label{eq:def_2point}
\left< \Psi_a(k_1, \eta) \Psi_b(k_2, \eta) \right>
\equiv \delta^{(3)}(k_1 + k_2) \, P_{ab}(k_1,\eta). 
\ee
The conservation of momentum in the mode coupling term ensures that the 
two point correlator (\ref{eq:def_2point}) is proportional to the delta function
at all times. This reflects the translation invariance of the system.
The leading order contribution to 
it in SPT is the so-called linear power spectrum
\bea 
\label{eq:IPS}
P^L_{ab}(\eta) &\equiv& g_{ac}(\eta, \eta_0) g_{bd}(\eta, \eta_0) P^0_{cd}(k) \, ,
\eea
where we have introduced the initial power spectrum
\be
\left< \Psi_a(k_1, \eta_0) \Psi_b(k_2, \eta_0) \right>
\equiv \delta^{(3)}(k_1 + k_2) \, P_{ab}^0(k_1) \, .
\ee
Thus,
\bea
\label{eq:SPTPS}
P_{ab}(k, \eta) &=& P^L_{ab}(\eta)  + \cdots .
\eea
In the following we only deal with
Gaussian initial conditions that are uniquely specified by the initial power
spectrum $P^0(k)$.

\begin{figure}[t!]
\begin{center}
\includegraphics[width=0.95\textwidth, clip ]{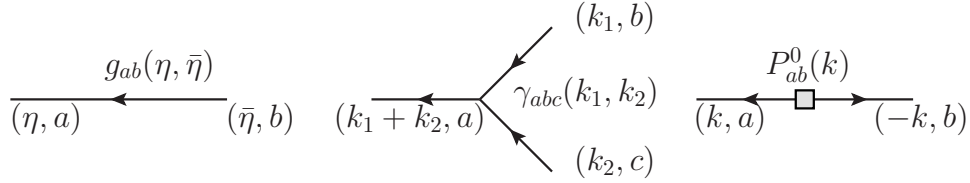}
\end{center}
\caption{
\label{fig:FeynmanRules}
\small Building blocks  of the Feynman rules of standard perturbation theory. }
\end{figure}

\begin{figure}[t!]
\begin{center}
\includegraphics[width=1.00\textwidth, clip ]{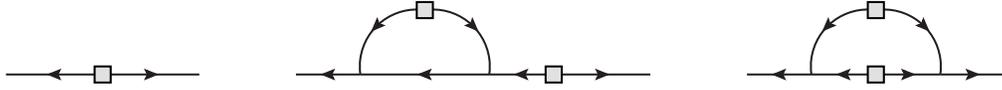}
\end{center}
\caption{
\label{fig:PS_oneloop}
\small All tree level and one-loop contributions to the power spectrum. }
\end{figure}

The perturbative expansion can be pictorially represented by classical Feynman
rules \cite{Crocce:2005xy}. The two building blocks are the vertex and the 
Green's function displayed in
Fig.~\ref{fig:FeynmanRules}. The causal structure of the integral
(\ref{eq:Psi_int}) enforces a flow of time that is indicated by the
arrow. The initial power spectrum is indicated by a box. 
 The leading topologies
that contribute to the power spectrum are shown in
Fig.~\ref{fig:PS_oneloop}.

\section{Renormalized perturbation theory and the eikonal approximation\label{RPT_eik}}

\begin{figure}[t!]
\begin{center}
\includegraphics[width=0.4\textwidth, clip ]{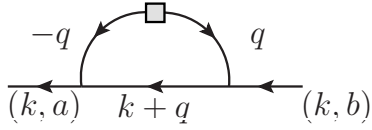}
\end{center}
\caption{
\label{fig:oneloop_prop}
\small One-loop contribution to the propagator. }
\end{figure}

In this section we briefly review the key aspects of renormalized
cosmological perturbations (RPT)~\cite{Crocce:2005xy,Crocce:2005xz}
(see also \cite{Wyld61,LP95} for similar resummations in the case of fluid dynamics)
and the eikonal approach~\cite{Bernardeau:2011vy} which constitute the
basis for our later discussions. Both schemes
are motivated to address convergence issues
related to the enhancement of the vertex 
(coupling) between hard and soft modes. 
For example, the diagram in Fig.~\ref{fig:oneloop_prop}
scales for $k \gg q$ as 
\be
\label{eq:soft_corr}
g_{ab}(\eta, \eta_0) \,
\int_{\eta_0}^\eta d\bar \eta \int_{\eta_0}^{\bar \eta} d\tilde \eta \, g_{2c}(\bar \eta, \eta_0) g_{2d}(\tilde
\eta, \eta_0)
 \int d^3 q \frac{(k \cdot q)^2}{q^4}  P_{cd}^0(q)  \, . 
\ee
So even if the final integral is finite and the growing of the Green's function 
 small, there results an enhancement from attaching
more soft loops to a line with momentum $k$ if $k$ is large enough, i.e. if $k \, \sigma_d \gg 1$,
where
\be
\label{eq:sigma}
\sigma_d^2 (\Lambda,\eta)\equiv \frac{1}{k^2} \int^\Lambda d^3q \frac{(k \cdot q)^2}{q^4}\, {P^L (q,\eta)}
= \frac{4\pi}{3} \int^\Lambda d q\, {P^L (q,\eta)} \,, 
\ee
where  $P^L\equiv P^L_{22}$ and $\Lambda$ represents the scale at which we cut-off the initial power spectrum.
This is precisely the scale of non-linearity discussed in the 
introduction, Eq.~(\ref{eq:kNL}). In particular, $k^2 \sigma_d^2$ 
can be larger than the dimensionless variance of the density field \cite{Scoccimarro:1995if}
\be
\label{eq:NLFS}
\sigma^2_l(\Lambda,\eta)\equiv 4\pi \int^\Lambda dq\,q^2 P^L(q,\eta), 
\ee
which would control the non-linear scale in the absence of vertex enhancement\footnote{We 
chose the notation for these two quantities that best adapts to previous literature. We also define $\sigma_{d,l}(\Lambda)\equiv\sigma_{d,l}(\Lambda,\eta=0)$ for the values today, and $\sigma_d\equiv\sigma_d(\Lambda\to\infty)$; note that this limit exists for realistic power spectra. In contrast, $\sigma_l(\Lambda)$ has a logarithmic sensitivity to $\Lambda$, as will be discussed in detail in section~\ref{sec:discussion}.}. 
At this stage it seems that to  describe physics
beyond the scale $\sigma_d$ one should
first be able to sum up all soft contributions to the propagation 
of hard modes. 
Once this is done, one expects
to find a perturbation theory with better convergence
properties.
This idea was put on a firm basis in the theory of RPT \cite{Crocce:2005xy,Crocce:2005xz}.
In this case,  the role of the Green's function (\ref{eq:Green}) in the Feynman rules
 is played by the \emph{propagator} 
\be
\label{eq:def:full_prop}
G^{(1)}_{ab}(k, \eta, \bar \eta) \delta^{(3)}(k - k_1) 
\equiv \left< \frac{\delta \Psi_a(k, \eta)}{\delta \Psi_b(k_1, \bar \eta)} \right> \, ,
\ee
that in terms of perturbation theory contains all diagrams with exactly one
incoming and one outgoing line and arbitrary (soft or not) loop corrections to
it~\footnote{The meaning of the superscript $(1)$ is clarified below.}. 

As we will review in the next section, the leading behavior for
large momenta $k$ (and considering only the contribution from the
growing mode)
 can be resummed for the propagator of renormalized perturbation
theory~\cite{Crocce:2005xy} 
\be
\label{eq:GRPT}
G_{RPT}^{(1)} (k, \eta, \bar \eta) = g_{ab} (\eta, \bar \eta) \exp \lp -\frac12
k^2 \sigma_d^2(a(\eta)-a(\bar \eta))^2 \rp,
\ee
where we used $a(\eta)\simeq e^\eta$ and set $a=1$ today.

The perturbative expansion in RPT can be 
formulated by means of the 
 $n$-point propagators defined as \cite{Bernardeau:2008fa}
\be
G^{(n)}_{a{a_i} \dots {a_n}}(k_i, \eta, \bar \eta) \delta^{(3)}(k-\sum k_i) \equiv \frac{1}{n!}
\left< \frac{\delta^n \Psi_a(k,\eta)}{\delta\Psi_{a_1}(k_1, \bar\eta) \cdots \delta\Psi_{a_n}(k_n,
\bar\eta)} \right> \,.
\ee
In terms of those,  the power spectrum at late times can be written as
\bea
\label{eq:power_resummed}
P_{ab}(k, \eta) &=& \sum_n \, n! \, \left( \prod_{i=1}^n \int d^3k_i \, 
P_{{a_i}{b_i}}^0(k_i) \right) \, \delta^{(3)}(k - \sum k_i )\,  \nn \\
&& \, \times \, G_{a{a_1}\dots{a_n}}^{(n)} (k_i,\eta, \eta_0)
G_{b{b_1}\dots{b_n}}^{(n)} ( -k_i,\eta, \eta_0) \, .
\eea
Pictorially, $G^{(n)}$ is the sum of all diagrams with $n$ incoming lines and
one outgoing
line. 
As long as the resummed $n$-point functions $G^{(n)}$ are well behaved for large
$k$, the sum should converge. It was shown in \cite{Bernardeau:2008fa}
that the corrections
from soft modes to the individual $n$-point propagators
also generate an exponential suppression related to the scale $\sigma_d$.

For the power spectrum and related observables, the previous results
provide only a partial resummation of the effects of soft modes. This is
commonly encoded in an expression  of the form (compare with 
(\ref{eq:SPTPS}))
\be
\label{eq:resPS}
P= \left(G_{RPT}^{(1)} \right)^2P^0 +P_{MC},
\ee 
where the piece $P_{MC}$ contains all the contributions from other diagrams.
From the previous arguments, one may expect  $P_{MC}$ to be
 better behaved than the corrections in Eq.~(\ref{eq:SPTPS}) and that
RPT with the leading effect of the soft modes to the propagator resummed
have better convergence  properties than SPT.

To check this, one needs to deal with \emph{all} soft corrections to the hard skeletons behind observables. 
This is the purpose of the eikonal approximation~\cite{Bernardeau:2011vy}. The
main observation of this approach is that in (\ref{eq:Psi_eta}) the contribution from the soft
modes to the mode coupling is approximated by
\bea
 && \hskip -2 cm \int d^3k_1 d^3k_2 \, \gamma_{abc}(k_1,k_2) \Psi_b(k_1, \eta)
\Psi_c(k_2, \eta)  \delta^{(3)}(k - k_1 - k_2)\nonumber \\
 &&\hspace{3cm} \simeq   \Psi_a(k, \eta) \, \int d^3q \, \frac{k \cdot q}{q^2} \Psi_2(q,
\eta) \, .
\eea
Using this relation in (\ref{eq:Psi_eta})  leads to the result (cf. (\ref{eq:PsiL}))
\be
\label{eq:Psi_eikonal}
\Psi (k, \eta) = g(\eta, \eta_0) \exp 
\left[ \int_{\eta_0}^\eta d \tilde \eta \int d^3q  \frac{k \cdot q}{q^2} \Psi_2(q,\tilde \eta) 
\right]
\Psi (k,\eta_0) \, .
\ee
In turn, for the propagator (\ref{eq:def:full_prop}) in the eikonal
approximation this yields
\be
G^{(1)}_{\rm eikonal} ( k,\eta, \bar \eta) = g(\eta, \bar \eta) \left< 
\exp  \left[ \int_{\bar\eta}^\eta d \tilde \eta\int d^3q  \frac{k \cdot q}{q^2} \Psi_2(q,\tilde \eta)
 \right]\right> \, ,
\ee
that for Gaussian initial conditions is (in leading order, see (\ref{eq:PsiL})) given by the second cumulant
\be
\label{eq:Geik}
G^{(1)}_{\rm eikonal} ( k,\eta, \bar \eta) = g(\eta, \bar\eta)  
\exp \left( \frac{c_2}{2} \right) \, ,
\ee
with
\be
\begin{split}
c_2 \equiv &\left\langle\left( \int_{\bar\eta}^\eta d \tilde \eta\int d^3q  \frac{k \cdot q}{q^2}
 \Psi_2(q, \tilde \eta)\right)^2\right\rangle_{connected}\\
\hspace{1cm}& \simeq -\int_{\bar\eta}^{\eta} \, d\hat\eta \, d\tilde\eta \, g_{2a}(\hat
\eta,\eta_0) g_{2b}(\tilde \eta,\eta_0) 
\int d^3 q \frac{(k \cdot q)^2}{q^4} P_{ab}^0(q) \, .
\end{split}
\ee
To compare it  with (\ref{eq:GRPT}), we
consider only the growing mode, for which
\be
c_2=
-k^2 \sigma_d^2(a(\eta)-a(\bar \eta))^2,
\ee
which shows that  the eikonal
approximation
indeed resums the leading soft corrections. 
A possible physical interpretation of this propagator is that the modes of hard momentum $k$
are scattered by the  background of soft modes. This decorrelates the modes over time what leads
to the exponential fall-off of the propagator for late times. 

One of the advantages of the eikonal approximation is that it allows to go beyond the 
propagator and compute the effects of the soft modes in the 
power spectrum~\cite{Bernardeau:2011vy}. Since the exponent in 
 (\ref{eq:Psi_eikonal}) is odd under a sign flip of
$k$ and the power spectrum involves the combination $\langle\Psi(k)
\Psi(k')\rangle \propto \delta(k+k')$, the two exponential factors cancel and the resulting power
spectrum is unsuppressed. Hence, the complete resummation of the leading effect of
 soft modes
does not produce an expression as (\ref{eq:resPS}) (see also  (\ref{eq:power_resummed})),
but leaves the power spectrum (\ref{eq:SPTPS}) unchanged \cite{Bernardeau:2011vy}. 
This result can also be derived in the diagrammatic language of RPT, which we do in 
the next section.

\section{Resummation of general skeleton diagrams\label{skeletons_resummed}}

\begin{figure}[t!]
\begin{center}
\includegraphics[width=0.95\textwidth, clip ]{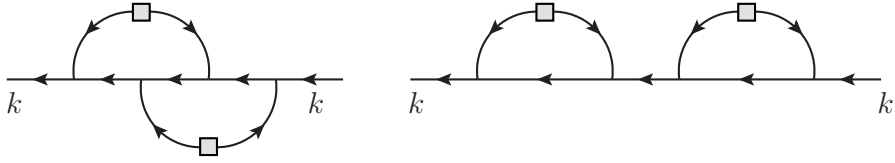}
\end{center}
\caption{
\label{fig:twoloop_prop}
\small Two-loop contributions to the propagator. }
\end{figure}

\begin{figure}[t!]
\begin{center}
\includegraphics[width=0.55\textwidth, clip ]{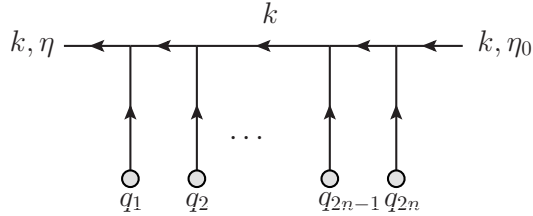}
\end{center}
\caption{
\label{fig:soft_ex}
\small  Soft modes attached to a hard linearized propagator. If the inflow of momentum is neglected, the order of the vertices is irrelevant. }
\end{figure}

Let us start this section by reviewing the resummation of the leading
contributions to the propagator as first given in~\cite{Crocce:2005xy}. We want
to resum all leading soft corrections in the limit of large $k$. A vertex with
one soft and one hard incoming mode is enhanced by a factor $k \cdot q/q^2$,
while a vertex with two similar modes is not enhanced. Hence, for fixed loop
order the dominant contribution comes from attaching all soft modes directly to
the hard mode that flows through the linearized propagator. Possible diagrams at one and
two loop order are shown in Figs.~\ref{fig:oneloop_prop}
and~\ref{fig:twoloop_prop}.
In the limit of soft corrections, the vertex is proportional to the Kronecker
delta (\ref{eq:soft_vertex}). The inflowing momentum can be neglected at leading order and
the Green's functions  $g(\eta_{i+1}, \eta_i)$
along the hard mode combine to one linearized propagator involving only the very first and
the very last time, $g(\eta, \eta_0)$. This is a generic result that  hinges on the group structure of the propagators.
In particular it is valid for any  expansion history of the Universe and beyond the restriction to the growing mode case. A diagram with $n$ loops involves attaching $2n$ linearized propagators of the form depicted in Fig.~\ref{fig:soft_ex}. 
As long as the inflow of soft momentum is neglected, the order of all the vertices is irrelevant.
The time integrations involves (with $\eta_{2n+1}\equiv\eta$)
\be
\prod_{i=0}^{2n} \int_{\eta_0}^{\eta_{i+1}} d\eta_i \,g_{2a}(\eta_i,\eta_0)\, .
\ee
Since we are considering the effect of all possible contractions with the initial power spectrum, 
the symmetries of the resulting expressions  allow us to rewrite the time
integrations as 
\be
\frac{1}{(2n)!} \prod_{i=0}^{2n} \int_{\eta_0}^{\eta} d\eta_i \, g_{2a}(\eta_i,\eta_0).
\ee
In total, there are $(2n-1)!!$ possibilities to contract the soft modes with initial power spectra.
Thus, at $n$-loop order one finds a contribution $(c_2/2)^n/n!$ and summing over loops
the eikonal result (\ref{eq:Geik}) is recovered. Notice that this result is valid for any
matrix $\Omega$, which just changes the form of the linearized propagator
and hence the particular value of $\sigma_d$. 

\subsection{Power Spectrum}

\begin{figure}[t!]
\begin{center}
\includegraphics[width=0.70\textwidth, clip ]{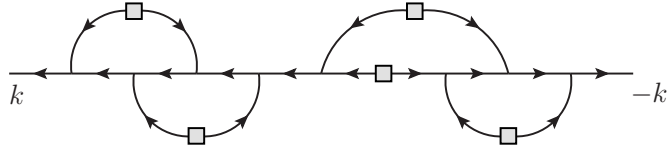}
\end{center}
\caption{
\label{fig:PS_xloop}
\small A contribution to the power spectrum at four-loop order
with $n_{ll}=2$, $n_{lr}=1$ and $n_{rr}=1$. }
\end{figure}

Next, we consider all soft corrections to the power spectrum. A generic graph at
four-loop order is shown in Fig.~\ref{fig:PS_xloop}.
Imagine there are $n_{ll}$ loop corrections to the left hard linearized propagator and
$n_{rr}$ loop corrections to the right linearized hard propagator. In addition, we denote
the number of soft loop corrections connecting the left linearized hard propagator with the
right one by $n_{lr}$. Following the same logic that applied for the propagator,
 one can extend all time integrations from $\eta_0$
to $\eta$ what leads to a factor $1/(2 n_{ll} + n_{lr})!$ for the left linearized hard
propagator and $1/(2 n_{rr} + n_{lr})!$ for the right one. There are $\binom{2
n_{ll} + n_{lr}}{n_{lr}}$ combinations to split the soft modes connected to the
left branch into the two groups, such that the combinatorial factor before
contraction is $1/(2 n_{ll})! n_{lr}!$ and similarly for the right soft modes
one finds $1/(2 n_{rr})! n_{lr}!$. As before, there are $(2n_{ll}-1)!!$
possibilities to contract the left loops and $(2n_{rr}-1)!!$ to contract the
right ones. In addition there are $n_{lr}!$ combinations to contract the left
modes with the right ones. Finally, notice that the left and right loops produce
a factor $c_2$ while the soft loops connecting the left branch with the right
one leads to a factor $-c_2$. In summary, one finds that the leading
corrections from soft modes for the power spectrum of a hard mode reduce
to the factor
\be
\label{eq:cancel}
\sum_{n_{ll}, n_{lr}, n_{rr}} \frac{1}{ n_{ll}! n_{lr}! n_{rr}!} 
\left(\frac{c_2}{2} \right)^{n_{ll}}
\left(\frac{c_2}{2} \right)^{n_{rr}}
(-c_2)^{n_{lr}}
= e^{c_2/2 - c_2 +c_2/2} = 1 \, .
\ee
Thus, the corrections that connect the different branches exactly
compensate for the exponential suppression from resumming the soft
corrections to the propagators. This is in complete accord with the eikonal
approximation that predicts a suppression in the propagator but not in the power
spectrum. 
This cancellation of the leading soft corrections to
the power spectrum was  first 
observed in \cite{Jain:1995kx}.
Here we gave a diagrammatic derivation (cf. also \cite{Anselmi:2012cn,Peloso:2013zw}).

The previous result implies that the expansion (\ref{eq:power_resummed}) does
not necessarily improve the convergence in the power spectrum compared to
standard perturbation theory (see \cite{Sugiyama:2013pwa} for similar
conclusions in the context of RegPT). According to the above resummation, in the 
limit where all $k_i$ but one momentum are soft the $n$-point functions
$G^{(n)}$ are enhanced such that 
 the sum produces an exponential that cancels the exponential suppression in
each of the $G^{(n)}$ observed in RPT \cite{Bernardeau:2008fa}.
 Parametrically,  the leading soft corrections to the different contributions to the sum (\ref{eq:power_resummed}) with resummed $n$-point propagators scale as $P \sim \sum_n \frac{1}{n!} (k\,  \sigma_d(\eta))^{2n} \exp(- k^2 \sigma_d(\eta)^2)$.  
So for large $k^2 \sigma_d(\eta)^2 \gg 1$, the sum can only start converging after $n
\simeq k^2 \sigma_d(\eta)^2$ terms are 
taken into account. In fact the situation is even worse, since reproducing the
leading $k$-behavior order by order 
is not enough to ensure the cancellation of subleading terms in the regime of large $k$ as we show below.

\subsection{General Skeleton}

The soft corrections to arbitrary skeletons of hard modes can also be 
resummed. For example
consider any of the one-loop contributions to the power spectrum depicted in
Fig.~\ref{fig:PS_oneloop} in a regime where none of the involved momenta is
soft. The diagram involves several linearized propagators with momenta $k_i$. Soft loops
can connect arbitrary linearized propagators and we denote the according number as
$n_{ij}$. The total number of soft lines at a linearized propagator $i$ is then $N_i =
2n_{ii} + \sum_{j\not=i} n_{ij}$. As before, the soft lines are time ordered  and
extending the time integrations over the full range leads to a factor $1/N_i!$.
The full range is hereby given by the time the original linearized propagator in the
skeleton diagram depends on that we denote $\eta_i$ and $\bar \eta_i$. Splitting
the soft lines into groups yields several binomial factors that 
contribute factors of the form $1/(2n_{ii})!/\prod_{j\not=i} n_{ij}!$. This is valid
for an arbitrary skeleton of hard modes.

Connecting a soft line of a linearized propagator $i$ with a soft line of a linearized propagator $j$
gives the integral
\be
- \int d^3 q \frac{k_i \cdot q}{q^2} \frac{k_j \cdot q}{q^2} P_{ab}^0(q) \, ,
\ee
and the time dependence\footnote{ Note that, since time has to increase along the arrows
of each linearized propagator, cf. (\ref{eq:Green}), the time $\tilde\eta$ of a soft vertex that is attached between
vertices at $\bar\eta_i$ and $\eta_i$ has to lie within the time interval $\bar\eta_i<\tilde\eta<\eta_i$.} 
\be
\int_{\bar \eta_i}^{\eta_i} d\tilde \eta \, g_{2a} (\tilde \eta, \eta_0) \,
\int_{\bar \eta_j}^{\eta_j} d\hat \eta \, g_{2b} (\hat \eta, \eta_0) \,
.
\ee
With the corresponding combinatorial factor to connect soft lines, the final
correction to the diagram  reads
\be
\label{eq:EikoY}
\exp \left[ - \frac12 \int d^3q Y_{2a} Y_{2b} P^0_{ab}(q) \right] \, ,
\ee
with
\be
\label{eq:Yab}
Y_{ab} \equiv \sum_i \int_{\bar \eta_i}^{\eta_i} d\tilde \eta \,  
\frac{k_i \cdot q}{q^2} g_{ab}(\tilde \eta, \eta_0) \, .
\ee
This result is actually expected from the eikonal approximation as we will see
in the next section. 

\begin{figure}[t!]
\begin{center}
\includegraphics[width=0.40\textwidth, clip ]{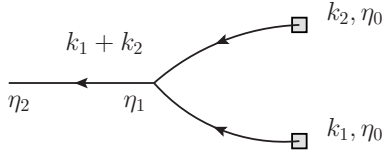}
\end{center}
\caption{
\label{fig:combining}
\small An example of several contributions to $Y_{ab}$ that can be combined. }
\end{figure}

Depending on the skeleton diagram, the function $Y_{ab}$ can be further
simplified. Consider two linearized propagators that originate from two initial 
power spectra and
end in a common vertex. The two corresponding contributions to $Y_{ab}$ then
share the same range of time integration and can be combined to one contribution
that is proportional to the sum of the two momenta of the skeleton diagram.
Hence, this term can also be combined with the subsequent propagator. An
illustration of this is given in Fig.~\ref{fig:combining}.
This fragment of a diagram acquires through soft corrections the contribution
\bea
Y_{ab} &\ni& \int_{\eta_0}^{\eta_1} d\tilde \eta \frac{k_1 \cdot q}{q^2} g_{ab}
(\tilde \eta, \eta_0)
+ \int_{\eta_0}^{\eta_1} d\tilde \eta \frac{k_2 \cdot q}{q^2} g_{ab} (\tilde
\eta, \eta_0) \nonumber \\
 && + \int_{\eta_1}^{\eta_2} d\tilde \eta \frac{(k_1+k_2) \cdot q}{q^2} g_{ab}
(\tilde \eta, \eta_0) \nn\\
&=& \int_{\eta_0}^{\eta_2} d\tilde \eta \frac{(k_1+k_2) \cdot q}{q^2} g_{ab}
(\tilde \eta, \eta_0).
\eea
Remarkably, the functions $Y_{ab}$ vanish for any equal-time correlator as for example
the power spectrum and the bispectrum. 

We will see in the next section
that this will allow us to go beyond the leading order considered in
 previous analysis \cite{Jain:1995kx, Bernardeau:2011vy} 
and show that the effects of soft modes over hard modes $k$
produce
at most an enhancement of $\log k$ compared to the linear power
spectrum  any order in perturbation theory.

\section{Eikonal approximation - the higher orders\label{sec:eikonal}}

In this section we extend the previous results and show that also
subleading soft corrections cancel each other in equal-time correlators.
This is done by using the eikonal approximation~\cite{Jain:1995kx, Bernardeau:2011vy} as the
zeroth order and systematically expanding the evolution equations around the eikonal
limit.  As has been shown in
\cite{Bernardeau:2011vy}, the eikonal approximation can be represented on the level of the equations of motion
by introducing a filter function that separates hard from soft modes, and rewriting (\ref{eq:Psi_eta}) as
\bea
\partial_\eta \Psi_a(k,\eta) + \Omega_{ab} \Psi_b (k,\eta) = \nn \\
&& \hskip -4 cm 2 F_\epsilon(k_1, k) \tilde \gamma_{abc}(k_1, k_2) \,  
\Psi_b(k_1, \eta) \Psi_c(k, \eta)  + \{ \gamma_{abc}  \Psi_b \Psi_c \}_H \, ,
\eea
where we used the notation 
\be\label{eq:gammatilde}
\tilde \gamma_{abc}(k_1, k_2) \equiv \delta_{b2} \delta_{ac} \frac{k_1 \cdot (k_1+k_2)}{2
k_1^2} \,.
\ee
The function $F_\epsilon$ is a filter that distinguishes soft modes from hard modes, for
example the Heaviside step-function $F_\epsilon(k_1, k) = \Theta (\epsilon |k| - |k_1|)$. 
In the previous construction the term $\{\dots\}_H$ denotes the difference between the full interaction
term (\ref{eq:Psi_eta}) and the first term. Only the first term on
the right side becomes large in the limit $|k_1|\ll |k|$.  
The eikonal approximation is obtained by neglecting the term $\{\dots\}_H$. The remaining
term can then be resummed and absorbed into the Green's function
\bea
\label{eq:g_eikonal}
g_{eikonal}(k, \eta, \eta_0) &\equiv& g(\eta, \eta_0) \exp 
\left[ \int_{\eta_0}^\eta d\tilde\eta \int d^3q \, F_\epsilon(q,k) \frac{k \cdot q}{q^2} \Psi_2(q,
\tilde\eta)  \right]
 \, \nn \\
&\equiv& g(\eta, \eta_0) \xi (k, \eta, \eta_0).
\eea
To discuss the cancellation of subleading terms it is necessary to discuss the rest term $\{\dots\}_H$ in some detail.
To this aim we add and subtract several terms to the right-hand side of (\ref{eq:Psi_eta}), and write the rest term
in the form\footnote{Recall that a term $\delta^{(3)}(k_1+k_2-k)$ 
and integrations in $k_1$ and $k_2$ are present in all the non-linear terms.}
\bea
\label{eq:Psi_eta_eik}
&&\hspace{-.5cm} \{ \gamma_{abc}  \Psi_b \Psi_c \}_H  \equiv
 {}  2 F_\epsilon(k_1, k) \tilde \gamma_{abc}(k_1, k_2) \,  
\Psi_b(k_1, \eta) (\Psi_c(k_2, \eta) - \Psi_c(k, \eta)) \, \nn \\
&& \hspace{1cm} {} +  F_\epsilon(k_1, k)  \left( \gamma_{abc}(k_1,k_2)  -
\tilde\gamma_{abc}(k_1, k_2) \right ) 
\Psi_b(k_1, \eta) \Psi_c(k_2, \eta) \, \nn \\
&&\hspace{1cm}  {} +   F_\epsilon(k_2, k)  \left( \gamma_{abc}(k_1,k_2)  -
\tilde\gamma_{acb}(k_2, k_1) \right ) 
\Psi_b(k_1, \eta) \Psi_c(k_2, \eta) \, \nn \\
&&\hspace{1cm}  {}  + (1-F_\epsilon(k_1, k) -F_\epsilon(k_2, k)) \gamma_{abc}(k_1,k_2)  
\Psi_b(k_1, \eta) \Psi_c(k_2, \eta) \, . 
\eea
 To go beyond the leading corrections we use  the eikonal approximation as the zeroth order solution and treat each of the
remaining terms in (\ref{eq:Psi_eta_eik}) perturbatively as an effective vertex. The
particular splitting into these four terms ensures that each diagram constructed in this scheme is free of any (residual) enhancement from 
soft modes (this will be discussed in more detail below).

The calculation of averaged quantities at a given order 
involves mixed cumulants of the field
due to the exponentiation of soft fields in (\ref{eq:g_eikonal}).
At leading order in the soft background
field, these cumulants reproduce the $Y$ functions (\ref{eq:Yab}) 
appearing in (\ref{eq:EikoY})~\cite{Bernardeau:2011vy}.

The perturbative expansion defined out of (\ref{eq:Psi_eta_eik})
 with the propagator (\ref{eq:g_eikonal}) seems better behaved 
than the one of SPT. 
This is because it tames the effects related to the scale (\ref{eq:sigma}).
This advantage 
 would justify its use despite
 its higher degree of complexity coming from
 the proliferation of vertices\footnote{  In particular some of the
 terms that have been added and subtracted in \eqref{eq:Psi_eta_eik} are not translation invariant,
 which leads to an apparent violation of momentum conservation in individual vertices and a perturbative expansion where single contributions to the power spectrum
 are non-diagonal in momentum space (although the full theory still respects translation invariance).} 
 and the  appearance
 of mixed cumulants mentioned above. 
But the results from 
the previous section show that the leading effect in $\sigma_d$ cancels for
 equal-time correlators and suggest that this may be a spurious scale. 
We are now going to prove this for all subleading effects. 
Thus, for these observables the convergence of any 
 scheme based on resumming the effects related to (\ref{eq:sigma})
 cannot be better than SPT.

We support this claim by interpreting the split of (\ref{eq:Psi_eta_eik}) in
terms of diagrams. In any diagram, the putative enhancement 
at scales $k\, \sigma_d \gg 1$ is related to soft momenta 
dressing a hard skeleton with momenta of order $k$.
In the previous section, we showed that the leading effect
(which at $n$-loop order scale as $(k\, \sigma_d)^{2n}$) cancels. 
To prove the cancellation of the 
subleading terms down to order ${\cal O}(k^0)$, 
we need to use the
perturbation theory based on (\ref{eq:Psi_eta_eik}). 
In comparison to the eikonal limit this requires  to 
account for the momentum injected into the diagram
from the soft loops, for the full vertex compared to \eqref{eq:gammatilde}
and finally for the 
self-coupling of soft modes (i.e. soft modes coupled to other soft modes).
Notice first that the vertices coming from the last three terms in (\ref{eq:Psi_eta_eik})
do not produce any enhancement and can be treated perturbatively. 
 The self-coupling of the soft modes is
in (\ref{eq:Psi_eta_eik}) represented by the fact that we resum the
time-dependent contribution $\Psi (q, \eta)$ (the whole non-linear
form). Whenever this propagator is used for vertices that
conserve momentum, the phase will disappear.
Thus, the only place where the scale $\sigma_d$ may appear
is in the term that does not conserve momentum,
\be
\label{eq:diffterm}
2 F_\epsilon(k_1, k) \tilde \gamma_{abc}(k_1, k_2) \,  
\Psi_b(k_1, \eta) (\Psi_c(k_2, \eta) - \Psi_c(k, \eta)).
\ee
If $k_1 \ll k$, 
the vertex is enhanced but the contribution comes with a factor\footnote{This
expression is precise to order ${\cal O}(\epsilon)$.}
\bea
(\Psi (k_2, \eta) - \Psi(k,\eta)) &\simeq& g(\eta, \eta_0) \xi(k_2, \eta,
\eta_0) \nn \\
&& \times \left[ \Psi(k_2, \eta_0) - \Psi(k,\eta_0) \xi(k-k_2, \eta, \eta_0)
\right] \,  .
\eea
For equal-time correlators the factor $\xi(k_2, \eta, \eta_0)$ will combine with the remaining factors of the diagram to yield
unity, while the second term in brackets is not enhanced in $k$ any more and can
be treated perturbatively (recall the factor $F_\epsilon(k_1, k) $ in (\ref{eq:diffterm}))
\be
\xi(k-k_2, \eta, \eta_0) = \xi(k_1, \eta, \eta_0) \simeq  
1 +\int_{\eta_0}^\eta \int d^3q \, \frac{k_1 \cdot q}{q^2} \Psi_2(q, \eta) \, .
\ee
Hence, $(\Psi (k_2, \eta) - \Psi(k,\eta))$ is of order $k_1/k$, which cancels
the enhancement coming from $\tilde \gamma$, and the 
factors $\xi(k, \eta, \eta_0)$ also cancel in this vertex.

This proofs that the enhancement from soft modes 
characterized by the  scale $\sigma_d$ is completely absent in  equal-time  observables. 
This includes the matter power spectrum and any other $n$-point correlation function, like the bispectrum.
 Notice that the argument in this section is rather general. The main assumption is that the leading term in (\ref{eq:Psi_eta_eik}) can be treated non-perturbatively and that the resulting cumulants are not enhanced for large $k$. Another assumption is that the Gaussian random field $\Psi(k)$ can be expanded in a Taylor series, see also \cite{Jain:1995kx}. 
On the other hand, the precise form of the vertices is irrelevant for the argument to work
as far as  the full vertices respect translation invariance and provided that the soft-enhanced terms are proportional to the unit matrix $\tilde \gamma_{abc} \propto \delta_{ac}$ such that $\xi$ has no matrix structure. The latter condition is for example violated in a multi-component fluid for the non-adiabatic decaying isodensity modes~\cite{Bernardeau:2012aq}. Notice also that, since the proof does not rely on the form of the linear part  $\Omega$ (given by \eqref{eq:EdSOm} for Einstein de-Sitter), it is true for any cosmological expansion history, including any possible term that modifies the linear term. This means that it is valid as well for the Zel'dovich approximation
(\ref{eq:OZA}), in accordance with \cite{Scoccimarro:1995if}.

In the next section we complement the general proof presented above by an explicit check of the cancellation
for the matter power spectrum in an Einstein-de Sitter cosmology
at two-loop order, and for the loop integrand up to four-loop order.

\section{Matter power spectrum\label{power_spec}}

In this section we want to clarify the consequences of the cancellation
of subleading corrections related to the scale $\sigma_d$
for the matter power spectrum.
The absence of physical effects related to the scale $\sigma_d$
was already derived at leading order and all loops 
for the matter power spectrum in \cite{Jain:1995kx}. This was interpreted in \cite{Scoccimarro:1995if} as
a consequence of Galilean invariance  (see \cite{Peloso:2013zw} for a recent discussion).
In \cite{Scoccimarro:1995if,Fry:1993bj}
the explicit calculation for the matter power spectrum and bispectrum to two loops and
with scale invariant initial power spectrum was performed, and no enhancement was found (the corrections are related to the scale $\sigma_l$). Note that for scale-invariant initial spectra with a power-law index smaller than $-1$ the cancellation of soft corrections is closely related to the cancellation of infrared divergences  \cite{Scoccimarro:1995if}.

Even if our proof of section~\ref{sec:eikonal} for the absence of enhancement by soft modes is valid for any 
initial power spectrum, in this section we will work with a  realistic $\Lambda$CDM form. For this, an
analytic expression is given by  the prominent fit by Eisenstein and Hu~\cite{Eisenstein:1997jh}.
For analyzing the leading asymptotic behavior of loop corrections, it is often sufficient to
consider the restriction of this fit to the Einstein-de Sitter case with spectral index $n_s=1$,
\be
\label{eq:EH_spectrum}
P^0(k,\eta_0) \sim \alpha \frac{k \, L^2}{(L  + C \beta^2  k^2)^2} \, ,
\ee
with
\be
\label{eq:EH_parameters}
L \equiv \ln (e + 1.84 \beta k)\, , \quad
C \equiv 14.4 + \frac{325}{1 + 60.5 (\beta k)^{1.08}} \,, \quad \beta \simeq 6.86 \, \rm{Mpc}\,. 
\ee
For the normalization of the spectrum we fit the previous formula to the linear power spectrum 
at $z=0$ found in~\cite{Taruya:2012ut}, which yields $\alpha \simeq 17000 \, \rm{Mpc}^4$. Note that, for our numerical analysis, we use as input a linear power spectrum obtained with CAMB for WMAP5 parameters as provided in~\cite{Taruya:2012ut} (see the footnote \ref{fns} for the effect of considering $n_s\neq 1$).

\subsection{Formalism}

The different contributions to the power spectrum in
standard perturbation theory at $n$-loop order  can be written schematically as
\bea\label{eq:n-loop}
P_{n-loop}(k,\eta) &=& \sum_{m=1}^{n+1} \int d^3k_1 \, \dots d^3k_{n+1} \,
A^{(n)}_m(k_1, \dots , k_{n+1}) \, \nn \\
&& \times P^L(k_1,\eta) \dots P^L(k_{n+1},\eta) \, \delta^{(3)} (k-k_1\dots -k_m) \,.
\eea
For example at one-loop level, the two
diagrams in Fig.~\ref{fig:PS_oneloop}  involve the spectral dependence
$P^0(k)\, P^0(q)$ and $P^0(k-q)\, P^0(q)$,  which corresponds to $m=1$ and
$m=2$, respectively (as well as $q\equiv k_2$). Those diagrams
are  related to the terms denoted by\footnote{These quantities $P_{IJ}$
have nothing to do with the quantities $P_{ab}$ introduced above. 
It is always clear from the context to which quantity we refer.}
$P_{13}$ and $P_{22}$ in the standard perturbation theory \cite{Bernardeau:2001qr}. The expression
of $A^{(n)}_m$ in terms of the symmetrized kernels $F_n(k_1,\dots,k_n)$ that characterize
the non-linear evolution of the density field in the growing mode
in an Einstein-de Sitter Universe \cite{Bernardeau:2001qr,Fry:1993bj,Goroff:1986ep}
is shown in Eq.~(\ref{eq:Anm}). 

We are interested in the 
case where the momentum $k$ is much larger than the
 scale $\sigma_d^{-1}$. In this case, the previous integral seems to be dominated
by the regime in which all but one of the modes $k_i$ are soft.
This happens because the $A^{(n)}_m$ are
homogeneous rational functions in the momenta and in the previous regime
 the largest enhancement is of order $A^{(n)}_m \propto \prod_i (k\cdot k_i/k_i^2)^2 \propto |k|^{2n}$.
 Furthermore, 
the steep fall-off of the power spectrum  (\ref{eq:EH_spectrum}) at
high $k$ implies that the regime $k\,\sigma_d \gg \sigma_l(k)$
exists. Thus, we may concentrate on the non-linearities associated to $\sigma_d$.
 The results in the previous section imply that this is a naive expectation: there must be a cancellation between different
terms that eliminate this effect. 
To find this cancellation at different loop orders, we first
use the Dirac-delta function to perform the integration over $k_1$ in (\ref{eq:n-loop}). 
For large $k$, enhanced contributions from soft modes would originate from the domain $|k_i|\ll |k|$ for $i=2,..,n+1$
and $k_1\simeq k$ (possible ambiguities in choosing the hardest momentum to be $k_1$ are
taken into account by including appropriate Heaviside functions and combinatorial factors; see Appendix~\ref{sec:app:hard} for details). In order to isolate these terms, we Taylor expand 
\begin{equation}
P^L(k_1,\eta)\big|_{k_1=k-k_2\dots -k_m} =\sum_{l=0}b_l(k,q) [k\partial_k]^l P^{L}(k,\eta) , 
\end{equation}
where $q\equiv k_2+\dots+k_m$. The coefficients scale as $b_l\propto (|q|/|k|)^l$. Therefore, since the $A^{(n)}_m$
grow as $|k|^{2n}$, we need to expand up to $l\leq 2n$ to capture all terms that can potentially be affected by
enhancement for large $k$. By collecting all terms arising from the $l$-th term in the Taylor expansion, and relabelling the momenta, the $n$-loop power spectrum for large $k$ can be written in the form
\bea
\label{eq:P_asympt}
P_{n-loop}(k,\eta) &\to& \sum_{l \leq 2n} \int d^3k_1 \, \dots d^3k_{n} \,
B^{(n)}_l(k,k_1,
\dots , k_{n}) \nn \\
&& \times \left[ k \partial_k \right]^l P^L(k,\eta) P^L(k_1,\eta) \dots
P^L(k_{n},\eta) \, ,
\eea
up to terms that are suppressed by the hard external momentum, ${\cal O}(k_i/k)$.

The $B^{(n)}_l$ are linear combinations of the $A^{(n)}_m$ for $1\leq m\leq n+1$ multiplied by the appropriate coefficients $b_l$ and we provide explicit expressions in Appendix~\ref{sec:app:hard}, Eq.~\eqref{eq:Bnl}. Therefore, without any cancellations,
one would expect that $B^{(n)}_l \propto |k|^{2n-l}$ for large $k$. 
The results from \cite{Jain:1995kx} show that the leading contribution cancels, which means
that the enhancement is at most
$B^{(n)}_{l} \propto |k|^{2(n-1)-l}$. Our results of the previous section imply that this
cancellation is also valid for the subleading terms and that  
\be\label{eq:Bnl_largek}
B^{(n)}_l(k,k_1, \dots , k_{n}) \to C^{(n)}_l(k_1, \dots , k_{n}) + {\cal O}(k_i/k)\,,
\ee
where $B^{(n)}_l$ approaches a constant, denoted by $C^{(n)}_l$, for large $k$.

\subsection{Results\label{sec:results}}

\begin{figure}[t!]
\hspace*{-1.2cm}
\includegraphics[width=1.05\textwidth, clip ]{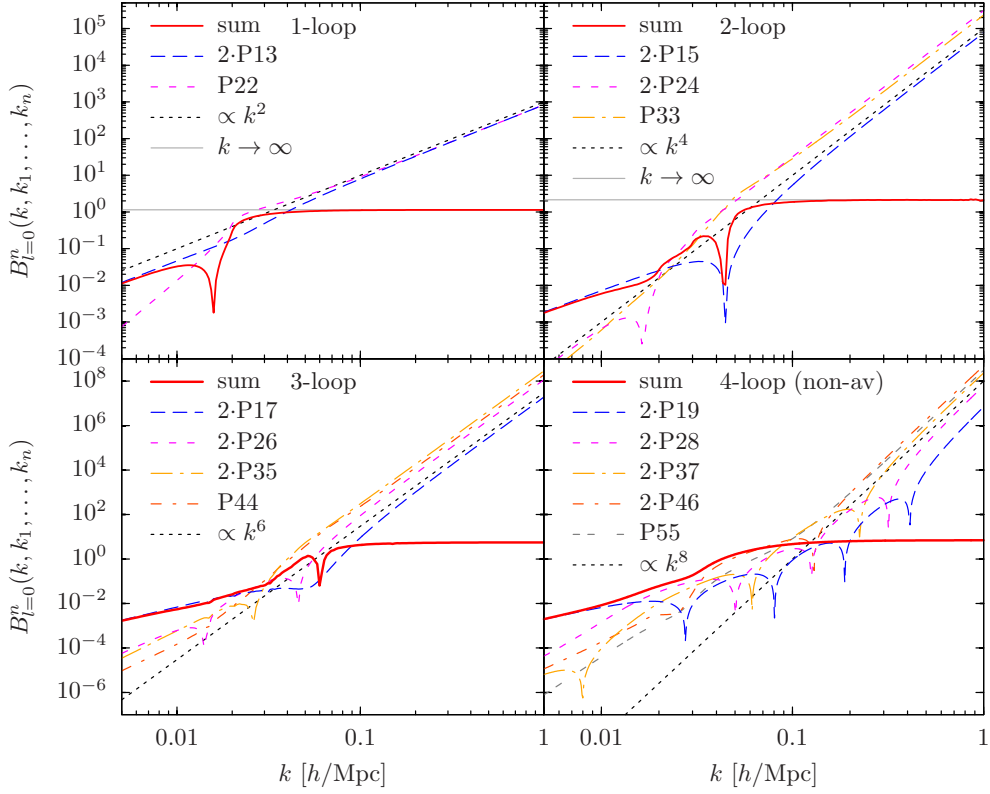}
\caption{
\label{fig:Bnl}
\small Cancellation of $n$-loop contributions to the integrand kernels $B^{(n)}_l$ of the power spectrum for $l=0$.
Shown are individual contributions $P_{IJ}$ (dashed lines) which grow like $k^{2n}$ (indicated by the black dotted lines). Their sum yields the kernels $B^{(n)}_l$ (red solid line), which approach a constant value as predicted in Eq.~\eqref{eq:Bnl_largek}. For comparison, the grey lines show the analytical results~\eqref{eq:B1llargek}, \eqref{eq:B2llargek} for the asymptotic value at one and two-loop. For the plot we have chosen $k_1=0.02, k_2=0.03, k_3=0.015, k_4=0.019$. For one-, two- and three-loop the kernels are averaged over the angles, see \eqref{eq:BnlAveraged}. The four-loop case corresponds to \eqref{eq:Bnl}.}
\end{figure}

\begin{figure}[t!]
\hspace*{-1.2cm}
\includegraphics[width=1.05\textwidth, clip ]{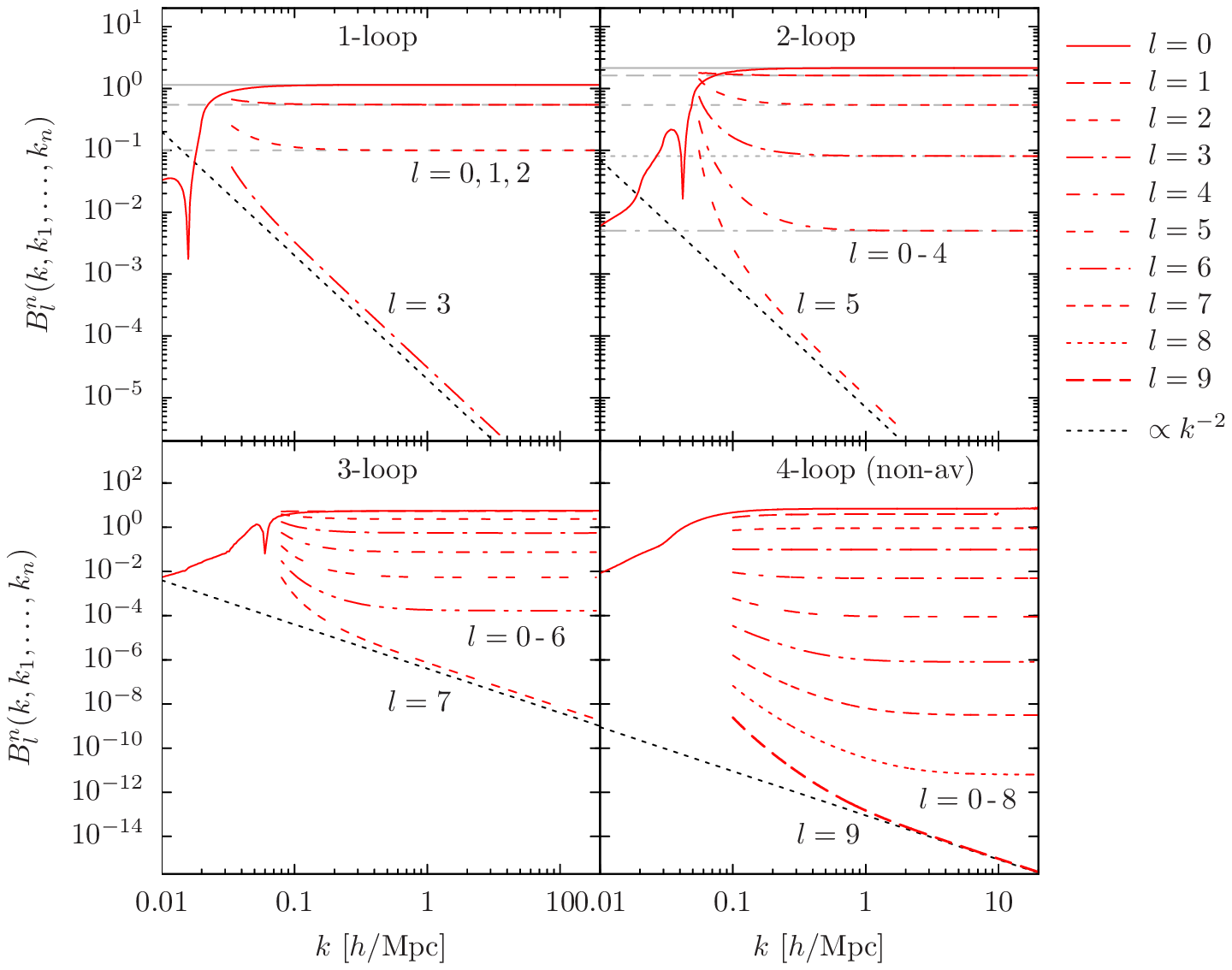}
\caption{
\label{fig:Bnl2}
\small Asymptotic behavior of $n$-loop integrand kernels $B^n_{l}$ of the power spectrum for $0\leq l \leq 2n+1$ (here only the sum of all individual contributions to each kernel is shown). The kernels with $l\leq 2n$ approach a constant value for large $k$ as predicted in Eq.~\eqref{eq:Bnl_largek}, and go to zero like $\mathcal{O}(k_i^2/k^2)$ for $l>2n$. Parameters are chosen as in Fig.~\ref{fig:Bnl}. The grey lines show the analytical large-$k$ results~\eqref{eq:B1llargek}, \eqref{eq:B2llargek} for comparison. For $l>0$ we show only the range where none of the Heaviside functions contained in the integrand \eqref{eq:n-loop-theta} is zero.}
\end{figure}

An explicit analytical calculation showing the cancellation of $k^4$- and subleading $k^2$-contributions  to the integrand kernels $B^{(n)}_l$ up to two loops is presented in the Appendix  \ref{sec:app:hard}. Explicit results for the asymptotic limit Eq.~\eqref{eq:Bnl_largek} are given in Eq.~\eqref{eq:B1llargek} and Eq.~\eqref{eq:B2llargek} for the one- and two-loop orders, respectively. In addition, we checked numerically for various momentum configurations that the cancellation of polynomially growing contributions and the asymptotic behavior Eq.~\eqref{eq:Bnl_largek} is indeed correct up to four loop order, see Figs.~\ref{fig:Bnl} and~\ref{fig:Bnl2}.


Using the analytical results Eq.~\eqref{eq:B1llargek} and Eq.~\eqref{eq:B2llargek} for the asymptotic values of the integrand kernels $B^{(n)}_l$ it is possible to derive approximate expressions\footnote{We do not write explicitly the dependence on $\eta$ in the rest of this section, since it is trivial to retrieve.}
 (up to $O(1/k^2)$ and subleading logarithmic corrections) for the power spectrum at large $k$,
\begin{eqnarray}
\label{eq:largekapprox}
 &&\hspace{-.6cm}P_{1-loop}(k) \sim 
\left(1.14P^L(k)-0.55 k \partial_kP^{L}(k) + 0.1 [k \partial_k]^2P^{L}(k)\right)
\sigma_l^2(k) \,,\nonumber\\
 &&\hspace{-.6cm}P_{2-loop}(k) \sim  \big(2.14 P^L(k)-1.62 k \partial_kP^{L}(k)
+0.55 [k \partial_k]^2P^{L}(k)\nonumber\\
&& {} -0.082[k \partial_k]^3P^{L}(k)+0.005 [k \partial_k]^4P^{L}(k)\big)\sigma_l^4(k)\;.
\end{eqnarray}
From these expressions, it is clear that the expansion parameter
at large $k$ is given by powers of $\sigma_l^2(k)$. 
We also see that even though all the functions $B^{(n)}_l$ are of order unity, a logarithmic
enhancement in the final power spectrum can arise depending on the precise shape
of the initial power spectrum. For  the spectrum \eqref{eq:EH_spectrum} one finds\footnote{\label{fns}For general spectral index $\sigma_l^2(k) \propto \ln^2(k)[ k^{n_s-1}-k_0^{n_s-1}]/(n_s-1)$ where $k_0\sim 0.02\, h/$Mpc. Note that for $k\ll k_* \equiv k_0 \exp(1/|n_s-1|)$, one may safely expand in $n_s-1$ to estimate the scaling with $k$. For example, $k_*\sim 10^9\,h/$Mpc for $n_s=0.96$. All numerical results are based on the WMAP5 spectrum~\cite{Taruya:2012ut} with $n_s=0.96$. For the analytical discussion of the limit $k\gg k_0$ it is thus legitimate to use \eqref{eq:EH_spectrum} as long as $k\ll k_*$, which is very well satisfied within the regime of interest in this work.} $\sigma_l^2(k) \propto \ln^3(k)$ for large $k$. This is discussed in
section~\ref{sec:discussion}.  Additional enhancement can potentially come from the fact that the derivatives $(k \partial_k)^n P^L(k)$ are larger than the power spectrum $P^L(k)$ itself.

Before moving to higher loops,  it is interesting to remind the behavior at small $k$ where the
 linear approximation is expected to work rather well. To
understand the size of
loop corrections parametrically, notice that in the limit $k\to 0$, the dominating contributions to the
$n$-loop corrections arise from the diagrams usually denoted by $P_{1,2n+1}$ and
scale as $\propto k^2P^L(k)$ (see Appendix
\ref{sec:app:soft}).  The behavior of this quantity is discussed in detail in~\cite{Bernardeau:2012ux}. We do not have much  to add to this discussion but we want to identify the expansion parameter in this regime and compare it to the large $k$ limit.
For the one- and two-loop expressions\footnote{In the
analytic calculations we assume a flat matter-dominated cosmology and only
consider the growing mode.}, we find
\begin{eqnarray}\label{smallk}
 P_{1-loop}(k) &\to&  -\frac{61}{105}k^2 P^L(k) \frac{4\pi}{3}\int_0^\infty dq
P^L(q) = -\frac{61}{105}k^2\sigma_d^2 P^L(k) \,, \nn \\
 P_{2-loop}(k) &\to&  -\frac{44764}{143325}k^2 P^L(k) \frac{4\pi}{3}\int_0^\infty
dq P^L(q)J(q)\,, 
\end{eqnarray}
where we introduced the (dimensionless) function
\begin{equation}
\label{eq:Jdef}
 J(q) = 4\pi\int_0^q dp \, p^2 g(p/q) P^L(p) \,.
\end{equation}
An explicit
form for $g(x)$ is given in Appendix \ref{sec:app:soft}  (see also \cite{Bernardeau:2012ux}). 
The relevant aspects are that it is a smooth function satisfying  $p^2 \, g(p/q)=q^2  \, g(q/p)$. This means
 that $g$ is almost constant as long as $p$ and $q$ are not of the same order and tends to zero for large argument. 
 Thus, it further cuts-off the integrals of the form (\ref{eq:Jdef}) at large value of the integration variable.
We find $g(0)=1.54$, $g(1)=1$ and $g(\infty)\to 0$.
A 
 good approximation for $J(q)$  in the regime of integration of (\ref{eq:Jdef}) is thus given by taking $g(p/q)\sim 1$, which means that $J(q) \simeq \sigma_l^2(q)$. Therefore,  the relative importance of the two-loop results with respect
to the one-loop case is again related to the scale $\sigma_l(q)$. Notice, however, 
that in the low-$k$ case this function is integrated over (cf. (\ref{eq:Jdef})).

The previous calculations confirming the results of section~\ref{sec:eikonal} 
can be extended to higher loops. Even if the analytical calculations are very cumbersome in this case, 
one can still find general arguments about the different behaviors that may be checked with numerical computations.
For the large $k$ regime,  as long as $k$ is larger
than the momenta 
one integrates over in (\ref{eq:P_asympt}), the results of  section~\ref{sec:eikonal}  imply a roughly constant $B_l^{(n)}$ in \eqref{eq:P_asympt}. If one of
the momenta $k_i$ becomes larger than $k$, additional polynomial 
suppression $\propto k^2/k_i^2$ arises in the functions $B_l^{(n)}$ what renders the integrals in internal momenta finite. Therefore, $k$ effectively acts as a UV cutoff for the $k_i$ integrations, and one can estimate the
integral to contain at $n$-loop contributions of order ($l \leq 2n$)
\be
P_{n-loop} \ni \left([ k \partial_k ]^l P^L(k) \right)\,\sigma_l^{2n}(k) \qquad \text{(large $k$ limit)} \, .
\ee
Subleading logarithms depend on how the integrals 
are precisely cut off and can give sizable corrections. 

Concerning the low $k$ case, from the symmetries of the integrand in (\ref{eq:LowkP}) and its behavior at large momenta, 
one expects the power spectrum to behave as
\be\label{eq:smallk}
P_{n-loop}(k) \propto  k^2 P^L(k) \int_0^\infty dq P^L(q) \,\sigma_l^{2n-2}(q)  \qquad \text{(small $k$ limit)} \,.
\ee
As in the previous case,  the important expansion parameter for perturbation theory is related to the quantity
$\sigma_l^2$.

\begin{figure}[t!]
\begin{center}
\includegraphics[width=0.90\textwidth, clip ]{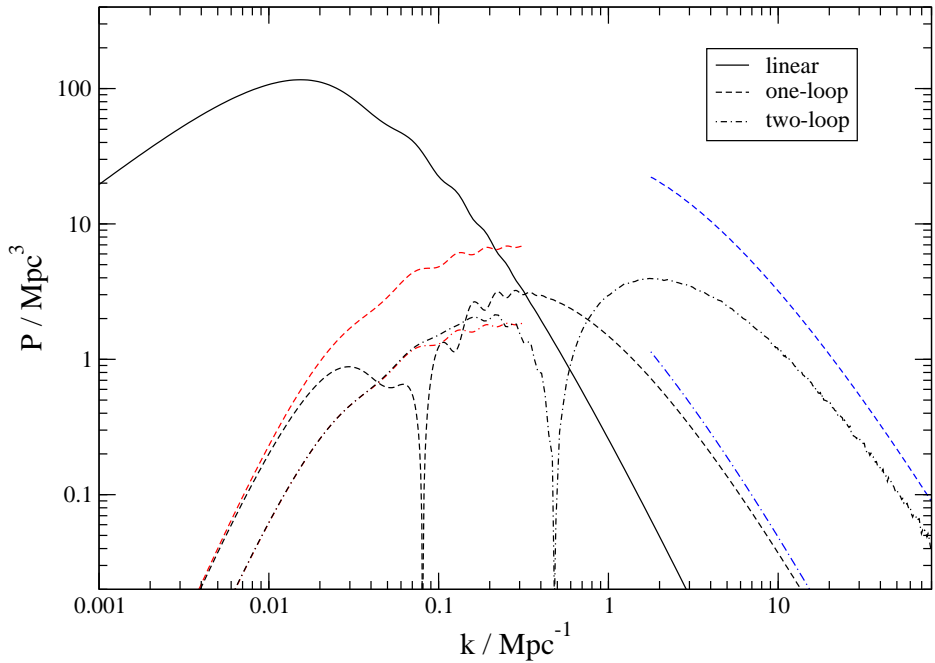}
\end{center}
\caption{
\label{fig:wmap5}
\small Linear power spectrum and the one- and two-loop corrections in
standard perturbation theory at $z=0$. The linear
spectrum corresponds to a $\Lambda CDM$ cosmology with WMAP5 parameters~\cite{Taruya:2012ut}. The blue
lines show the asymptotic behavior at large $k$ (see Eq.~(\ref{eq:largekapprox})), and the
red the one at small $k$ (see Eq.~(\ref{smallk})).}
\end{figure}

We have checked the previous asymptotic behavior by computing the power spectrum numerically up to two loops 
in an Einstein-de Sitter cosmology (taking into account only the growing mode). The results are displayed in Fig.~\ref{fig:wmap5}. 
We cross-checked our numerical results for the power spectrum with
the RegPT code~\cite{Taruya:2012ut} for momenta where the latter is available.
As can be seen in Fig.~\ref{fig:wmap5},
the asymptotic expressions (\ref{smallk}) and (\ref{eq:largekapprox}) agree with the full one- and two-loop results rather
accurately. The constant offset in the two-loop case at very large $k$ is expected because (\ref{eq:largekapprox}) captures only the leading logarithmic behavior. It is remarkable that even for very small
momenta, the two-loop contribution is only mildly suppressed compared to the
one-loop contribution. Naively, one might
expect a suppression by a power $k_0^2\sigma_d^2\sim\mathcal{O}(10^{-2})$, where $k_0$ is the position of the maximum of $P^L(k)$. However, the integral over $J(q)\sim \sigma_l^2(q)$ in (\ref{smallk}) is sensitive to the power
spectrum at smaller scales, and yields only a mild suppression with respect to
the one-loop correction. According to (\ref{eq:smallk}), the three-loop result should even
exceed the one-loop contribution at asymptotically small momenta for $z=0$. Actually, we checked numerically for a
few values of the momentum that this is indeed the case, and it is also in accordance with the findings of~\cite{Bernardeau:2012ux}. The same is true for the large $k$ regime. Here, the
logarithmic dependence in $\sigma_l^2(k)$ is clearly seen 
in the numerical results.

\section{Discussion\label{sec:discussion}}

Cosmological perturbation theory provides a very successful framework to understand the
gravitational clustering in the Universe, responsible for  its structure at the largest scales. 
Despite of this success, its range of validity (in the sense of convergence of the perturbative
series to the non-linear solution) is limited due to the growth of the importance of the non-linear contributions with time.
To devise methods that deal with these non-linear corrections it is crucial to understand which effects are behind the failure of perturbation theory. This is even more important when one realizes that a large amount of information about cosmological
evolution lies at scales close to the linear scales where these methods may be very useful. An example
is provided by the baryonic acoustic oscillations (BAO) at low redshift \cite{Crocce:2005xy,Eisenstein:2006nj,Weinberg:2012es}.

A first look at the structure of the standard formulation of cosmological perturbation theory (SPT)
singles out the functions
\bea
k^2\sigma_d^2(\Lambda,\eta) \equiv \frac{4\pi k^2}{3} \int^\Lambda dq {P^L (q,\eta)} , \hspace{.5cm}
\sigma_l^2(\Lambda,\eta) \equiv 4\pi \int^\Lambda dq \,q^2 P^L (q,\eta),
\eea
as responsible for the failure of the linearized approach for scale $k$, once one of them
becomes big. The first quantity, which is strongly $k$ dependent, is small 
at the peak of the power spectrum $k_0$,  $\sigma_d^2 k_0^2 \simeq 0.0135$ at $z=0$ and $k_0 \simeq 0.02 \, \rm{Mpc}^{-1}$ (see
\eqref{eq:EH_spectrum}). This number grows very fast with momenta, which is 
a consequence of the enhancement coming from the vertex $\gamma_{ijk}$
for a large hierarchy between the momenta of the two incoming modes, see (\ref{eq:soft_vertex}).
 At $n$-loop order, this can potentially lead to corrections to the power spectrum that scale at large $k$
as $\propto (k^2\sigma_d^2)^n P^L(k, \eta)$. It is well known that these leading soft corrections
cancel when summing over all $n$-loop diagrams \cite{Jain:1995kx}. However, there are also
subleading soft corrections, growing like $k^2$ at two-loop, like $k^4$ and $k^2$ at three-loop, etc.
 In our analysis we showed that,  when summing over all $n$-loop diagrams
all polynomially growing corrections $\propto k^{2m}$ with $1 \leq m \leq n$ cancel in the limit of large $k$.
This cancellation had not been proven
before to the best of our knowledge. Remarkably, the same cancellation happens
for any hard skeleton corresponding to $n$-point correlation functions evaluated at
equal-time.
While the general proof relies on the eikonal approximation,
we also checked explicitly that the subleading $k^2$-terms cancel at two-loop, and furthermore
checked numerically that all subleading $k^{2m}$-terms cancel up to four loop order.
The cancellation of the leading soft corrections was related 
to Galilean invariance in~\cite{Scoccimarro:1995if}. Physically, it seems plausible that similar
arguments could explain our results  (see e.g. \cite{Peloso:2013zw,Kehagias:2013yd}), though we are not aware of any explicit calculation that includes all subleading effects.

Our result also implies that in numerical calculations in SPT it is advantageous to sum over all relevant diagrams (and to symmetrize them appropriately, see the comments after Eq.~\eqref{eq:Bnl}) before any integration is
performed. In this way, the cancellation of different contributions occurs already on the level of the integrand and does not rely on the numerical accuracy of the integration.

 Even though the polynomially growing terms
cancel, there remains a logarithmic enhancement at large $k$ for the
matter power spectrum, see Eq.~(\ref{eq:P_asympt}).
We find that the leading logarithmic (LL) contributions at large $k$ are given by~\footnote{
See Eq.~(\ref{eq:largekapprox}) for the explicit expressions at one and two loops. }
\be
 P_{n-loop}^{LL}(k, \eta) \simeq \sum_{l=0}^{2n} c^{(n)}_l\; [k \partial_k]^l P^{L}(k, \eta)\; \sigma_l^{2n} (k, \eta)\,,
\ee
with some coefficients $c^{(n)}_l$ of order unity. 
This expression implies a growth with $k$ of the expansion parameter $\sigma_l^2 (k,\eta)$ which is logarithmic.
Note that  $\sigma_l^2 (k,\eta)$ also controls the loop expansion at the opposite limit of 
small $k$, see Eq.~(\ref{eq:smallk}).

The function $\sigma_l^2 (k,\eta)$ is not only sensitive to the high-momentum tail of the spectrum  (which makes it
increase logarithmically) 
but for the realistic case (\ref{eq:EH_spectrum}) it is also numerically rather large for small redshift $z\sim 0$, 
\be
\sigma_l^2(k,\eta) \simeq 0.15 \left(\frac{1}{1+z}\right)^2\, \big(\ln(e+1.84\beta k)\big)^3 \, .
\ee
We find that this function is the true expansion parameter of standard
perturbation theory for the equal-time power spectrum. For 
$k \sim 1 \, \rm{Mpc}^{-1}$ it is $\sigma_l^2(k,z=0) \sim 3.42$. This large value arises partially
due to the logarithmic dependence but also because the initial power spectrum (\ref{eq:EH_spectrum}) is parametrically 
enhanced by a factor $(C \beta^2 k_0^2)^{-2} \sim 16$. According to our arguments a smaller
value for $\sigma_l$ would improve the convergence
of linearized perturbation theory remarkably. We show this in Appendix~\ref{sec:app:conv} by using a fake initial power spectrum with the same $\sigma_d$ as $\Lambda$CDM but smaller $\sigma_l$. The results are shown in  Fig.~\ref{fig:spec_conv}, which should be compared with Fig.~\ref{fig:wmap5}.

Our conclusions about the matter power spectrum also hold for any correlation function
at equal times. In these observables, the effects related to the scale $\sigma_d$ are absent, and the departure
from the linear regime will be dominated by $\sigma_l$. The analysis is equally valid for arbitrary cosmologies, or even for departures
from the standard equations, as the Zel'dovich approximation.
 We would like to emphasize that this is not the case for other cosmological observables. 
For instance, the propagator as defined in (\ref{eq:def:full_prop}) is certainly affected by the enhancement of soft modes related to the scale $\sigma_d$. Its measurement
in simulations confirms this behavior \cite{Crocce:2005xz}, and it would be very interesting
 to extract it from real data. 
Furthermore, correlations between different redshift bins are also used for lensing tomography~\cite{Hu:1999ek}.

All resummation schemes in the literature resum only certain subsets of diagrams of SPT. 
Notice that those subsets do not necessarily reproduce the cancellation we 
found for equal-time correlations. Thus,
 the non-linearity associated with the scale $\sigma_d$ may be reintroduced as a
purely spurious effect.
We would like to emphasize that our results do not imply that standard
perturbation theory is superior to resummation schemes as e.g.~RPT. It might well be that at intermediate scales, these schemes
resum just the right subdiagrams to lead to accurate results  \cite{Bernardeau:2011dp}. In addition, they are very useful to describe
correlations at unequal times.
Still, our analysis supports that at large momenta resummation schemes that involve only the scale $\sigma_d$ cannot improve the
determination of the equal-time correlators systematically. 
We hope that our results are helpful to identify approximation schemes that respect the cancellation
of soft corrections and, at least partially, resum corrections related to the scale $\sigma_l$.

\section*{Acknowledgements}

We would like to thank Martin Kunz, Julien Lesgourgues, Valeria Pettorino, Antonio Riotto and Rom\'an Scoccimarro  for very useful discussions. DB would like to thank IPMU and DESY for their warm hospitality during the development of this work. This work has been partially supported by the German Science Foundation (DFG) within the Collaborative Research Center 676 ``Particles, Strings and the Early Universe''.

\begin{appendix}
\section{The power spectrum in the hard regime\label{sec:app:hard}}

In this appendix\footnote{To improve readability we will not write explicitly the
dependence on $\eta$ in the appendices, since it is trivial to retrieve.
In addition we display arrows on three vectors to facilitate discrimination
from absolute values. } 
we discuss the asymptotic behavior of loop corrections to the
power spectrum at large wavenumbers $k$ (small scales). It is possible to
rewrite the loop contributions such that the cancellation of polynomially
growing corrections $\propto k^n P^L(k)$, with $n>0$, is manifest. We first
demonstrate this for the one-loop corrections $P_{1-loop}=2P_{13}+P_{22}$, with
\begin{eqnarray}
 P_{13}(k) &=& 3 P^L(k) \int_q F_3^s(\vec k,\vec q,-\vec q) P^L(q) \,, \\
 P_{22}(k) &=& 2 \int_q\left[ F_2^s(\vec q,\vec k-\vec q)\right]^2 P^L(q) P^L(|\vec k - \vec
q|)\,, \nonumber
\end{eqnarray}
where $F_n^s(\vec q_1,\dots,\vec q_n)$ are the symmetrized PT kernels entering at
the different orders of SPT (see
e.g.~\cite{Bernardeau:2001qr}), $\int_q\equiv \int d^3q$ and $q\equiv|\vec q|$.
For large $k$ the two contributions asymptotically grow as $P_{13}\to
-k^2\sigma_d^2 P^L(k)/2$ and $P_{22}\to k^2\sigma^2/2 P^L(k)$, and it is evident
that the quadratically growing corrections cancel \cite{Vishniac}. At two-loop,
$P_{2-loop}=2P_{15}+2P_{24}+P_{33}$, the individual contributions
\begin{eqnarray}
 P_{15}(k) &=& 15 P^L(k) \int_{p,q} F_5^s(\vec k,\vec p,-\vec p, \vec q,-\vec q)
P^L(p)P^L(q) \,, \\
 P_{24}(k) &=& 12 \int_{p,q} F_2^s(\vec q,\vec k-\vec q) F_4^s(\vec q,\vec
k-\vec q,\vec p,-\vec p)P^L(p)P^L(q)P^L(|\vec k - \vec q|) \,,\nonumber\\
 P_{33}(k) &=& 9\int_{p,q}F_3^s(\vec k,\vec p,-\vec p)F_3^s(\vec k,\vec q,-\vec
q)P^L(p)P^L(q) \nonumber\\
 && {} + 6\int_{p,q} F_3^s(\vec p,\vec q,\vec k-\vec p-\vec q)^2
P^L(p)P^L(q)P^L(|\vec k -\vec p - \vec q|) \,, \nonumber
\end{eqnarray}
grow asymptotically as $k^4\sigma_d^4P^L(k)$. Again, it is straightforward to
check that the $k^4$ terms cancel each other.
However, there exist also subleading terms that grow as $k^2$. In order to show
that they also cancel in the
sum of all two-loop contributions, it is convenient to rearrange the various
terms. To demonstrate this, we first discuss
an analogous rearrangement for the one-loop terms,
\begin{equation}
 P_{1L}(k) = P^L(k) \int_q B^{(1)}_0(k,q)P^L(q) + \tilde P_{22}(k) \,,
\end{equation}
where $(d\Omega_q\equiv \sin\theta_q d\theta_q d\phi_q)$
\begin{eqnarray}
 B^{(1)}_0(k,q) &\equiv& \frac{1}{4\pi} \int d\Omega_q \left( 6 F_3^s(\vec
k,\vec q,-\vec q) + 4F_2^s(\vec q,\vec k-\vec q)^2 \right) \,,\\
 \tilde P_{22}(k) &\equiv& 4 \int_q F_2^s(\vec q,\vec k-\vec q)^2 P^L(q) \left(
\Theta(|\vec k - \vec q|-q)P^L(|\vec k - \vec q|) - P^L(k) \right)\,. \nonumber
\end{eqnarray}
In the latter contribution we subtracted a term $\propto P^L(k)$, such that the
difference in the bracket scales as $(\frac{\vec k\vec q}{k^2} +
\mathcal{O}(q^2/k^2)) k\partial_k P^L(k)$ for large $k$. The linear term
$\propto \vec q/k$ vanishes when integrating, such that the bracket leads to a
$q^2/k^2$ suppression that compensates the quadratic enhancement coming from the
kernel $(F_2^s)^2\sim (\vec k\vec q/q^2)^2/4$. Note that we also inserted a
Heaviside function compared to $P_{22}$ and
multiplied by two, which does not change $P_{22}$ due to the symmetry $\vec q
\to \vec k -\vec q$. Using the explicit expressions for the kernels, it is also
easy to check that quadratically growing terms $\sim k^2/q^2$ cancel in the
combination $B^{(1)}_0$ after averaging over angles (or alternatively
symmetrizing the integrand w.r.t.~$\vec q\leftrightarrow -\vec q$). The explicit
result can be easily calculated,
\begin{eqnarray}
\label{eq:B1llargek}
 B^{(1)}_0(k,q) &=& \frac{625}{882} - \frac{11k^4}{294q^4}
-\frac{523k^2}{1764q^2}-\frac{q^2}{12k^2} \nonumber\\
 &&  {} - \frac{k^2-q^2}{q^2}\left( \frac{11k^3}{1176q^3} + \frac{9k}{112q} +
\frac{8q}{49k} - \frac{q^3}{48k^3} \right)\ln\left(\frac{k-q}{k+q}\right)^2
\nonumber\\
 &\to& \frac{2519}{2205} + \mathcal{O}(q^2/k^2) \qquad \mbox{ for }
k\to\infty\,.
\end{eqnarray}
A similar rearrangement can be done for the two-loop contributions to the power
spectrum. In order to extract subleading $k^2$-terms apart from the leading
$k^4$, we have to re-shuffle terms proportional to the first and second
derivatives of the power spectrum ($\,{}'\equiv d/d\ln k$),
\begin{equation}
P^L(|\vec k - \vec q|) = P^L(k) +b_1(\vec k,\vec q)P^{L'}(k)  + b_2(\vec k,\vec
q)P^{L''}(k) + \dots \,,
\end{equation}
where 
\begin{equation}
 b_l(\vec k,\vec q) = \frac{1}{l!}\left[\ln\left(\frac{|\vec k-\vec q|}{|\vec
k|}\right)\right]^l\;.
\end{equation}
After this rearrangement, the two-loop contribution can be rewritten as
\begin{eqnarray}
\label{eq:P2L}
 P_{2L}(k) &=& P^L(k) \int_{p,q} B^{(2)}_0(k,p,q)P^L(q)P^L(p) \nn \\
&& {} + P^{L'}(k)
\int_{p,q} B^{(2)}_1(k,p,q)P^L(q)P^L(p)  \nonumber \\
 && {} + P^{L''}(k) \int_{p,q} B^{(2)}_2(k,p,q)P^L(q)P^L(p) + 2\tilde P_{24}(k) \nn \\
&& {}  +
\tilde P_{33}(k)\,,
\end{eqnarray}
where
\begin{eqnarray}
 B^{(2)}_l(k,p,q) &\equiv& \frac{1}{(4\pi)^2} \int d\Omega_q \int d\Omega_p
\Big( 30 F_5^s(\vec k,\vec p,-\vec p, \vec q,-\vec q) \delta_{l0} \nonumber\\
 && {} + 9 F_3^s(\vec k,\vec p,-\vec p)F_3^s(\vec k,\vec q,-\vec q)\delta_{l0}
\nonumber\\
 && {} + 18 F_3^s(\vec p,\vec q,\vec k-\vec p-\vec q)^2 b_l(\vec k,\vec p+\vec
q) \nonumber\\
 && {} + 24 F_2^s(\vec q,\vec k-\vec q) F_4^s(\vec q,\vec k-\vec q,\vec p,-\vec
p) b_l(\vec k,\vec q) \nonumber\\
 && {}  + 24 F_2^s(\vec p,\vec k-\vec p) F_4^s(\vec p,\vec k-\vec p,\vec q,-\vec
q) b_l(\vec k,\vec p) \Big)\,,
\end{eqnarray}
for $l=0,1,2$ with $b_0\equiv 1$, and
\begin{eqnarray}
 \tilde P_{24}(k) &\equiv& 24 \int_{p,q} F_2^s(\vec q,\vec k-\vec q) F_4^s(\vec
q,\vec k-\vec q,\vec p,-\vec p)P^L(p)P^L(q)\Big( \Theta(|\vec k - \vec q|-q)
\nonumber\\
 && {} \times P^L(|\vec k - \vec q|) - P^L(k) -b_1(\vec k,\vec q)P^{L'}(k)  -
b_2(\vec k,\vec q)P^{L''}(k)\Big)\,, \nonumber\\
 \tilde P_{33}(k) &\equiv& 18 \int_{p,q} F_3^s(\vec p,\vec q,\vec k-\vec p-\vec
q)^2P^L(p)P^L(q)\Big( \Theta(|\vec k -\vec p -\vec q|-q) \nonumber\\
 && {} \times \Theta(|\vec k -\vec p -\vec q|-p) P^L(|\vec k -\vec p - \vec q|)
- P^L(k) -b_1(\vec k,\vec p+\vec q)P^{L'}(k) \nonumber\\
 && {} - b_2(\vec k,\vec p+\vec q)P^{L''}(k)\Big)\,.
\end{eqnarray}
The subtracted terms inside the brackets in the last two expressions are
constructed such that, after angular integration, they compensate the
terms growing as $k^4$ and $k^2$ coming from the PT kernels for $k\to\infty$.
Thus, it remains to be shown that the functions $B^{(2)}_l$ in (\ref{eq:P2L}) do not grow with $k$. 
For that purpose, we expand the integrand for large $k$ and use that the angular
integration can equivalently be performed over $k$ and $p$ while keeping the direction
of $q$ fixed along the $z$-axis (see e.g. \cite{Scoccimarro:1995if}). We find that
indeed the $k^4$ and $k^2$ terms cancel. For the leading contributions in the
limit $k\to\infty$ (which are $\propto k^0$) we find
\be
\label{eq:B2llargek}
\begin{split}
 B^{(2)}_0(k,p,q) &\to \frac{271133}{92702610}\left( f(p/q) +
\frac{595172936}{813399} \right) + \mathcal{O}(1/k^2) \,, \\
 B^{(2)}_1(k,p,q) &\to \frac{5377}{12677280}\left(f(p/q)
-\frac{308888266}{80655}\right) + \mathcal{O}(1/k^2) \,,  \\
 B^{(2)}_2(k,p,q) &\to \frac{115}{32928}\left(-f(p/q) + \frac{452758}{2875}
\right) + \mathcal{O}(1/k^2) \,,\\
 B^{(2)}_3(k,p,q) &\to \frac{3}{3920}\left(f(p/q) - \frac{4882}{45} \right) +
\mathcal{O}(1/k^2) \,,\\
 B^{(2)}_4(k,p,q) &\to \frac{1}{200} + \mathcal{O}(1/k^2) \,,
 \end{split}
\ee
where
\begin{equation}
\label{eq:fx}
f(x) \equiv \frac{8x^2}{3} + \frac{8}{3x^2} - x^4 - \frac{1}{x^4} -
\frac{1}{4x^5} (x^2+1)(x^2-1)^4\ln\left(\frac{x-1}{x+1}\right)^2 \,.
\end{equation}
The results for $l=3,4$ are shown for later use, and
$B^{(2)}_l=\mathcal{O}(1/k^2)$ for $l\geq 5$. For completeness, we
also quote the large-$k$ limits for the higher-derivative contributions at
one-loop which are given by
\begin{equation}
 B^{(1)}_1(k,q) \to -\frac{23}{42}, \qquad
 B^{(1)}_2(k,q) \to \frac{1}{10}, \qquad
 B^{(1)}_{l\geq 3}(k,q) \to \mathcal{O}(1/k^2).
\end{equation}

The rearrangement presented above is useful to show that polynomially growing
corrections cancel out. The leading
correction for large $k$ is thus a logarithmic one. It is possible to extract an
analytic expression for the leading logarithmic corrections by performing a
similar rearrangement, but including terms up to $[ k \partial_k]^2 P^L$ at
one-loop and up to the fourth derivative $[ k \partial_k]^4 P^L$ at two-loop. This yields
approximate expressions for large $k$
\begin{eqnarray}
\label{eq:largek}
 P_{1-loop}(k) \hspace{-.2cm} &\to& \hspace{-.2cm} 
\left(\frac{2519}{2205}P^L(k)-\frac{23}{42}k \partial_kP^{L}(k)+\frac{1}{10}[k \partial_k]^2P^{L}(k)\right)
\sigma_l^2(k) \,,\\
\label{eq:largek2}
 P_{2-loop}(k)\hspace{-.2cm}  &\to&\hspace{-.2cm} (4\pi)^2\sum_{l=0}^4 [k \partial_k]^l P^L(k) \int_0^k dq q^2
P^L(q)\int_0^k dp p^2 P^L(p) B^{(2)}_l(k,p,q)  \nonumber\\
&& \hspace{-1.7cm}\sim \Bigg(\frac{1490372537}{695269575}P^L(k)-\frac{7719787}{4753980}k \partial_kP^{L}(k)
+\frac{112327}{205800}[k \partial_k]^2P^{L}(k)\nonumber\\
&& {} -\frac{1207}{14700}[k \partial_k]^3P^{L}(k)+\frac{1}{200}[k \partial_k]^4P^{L}(k)\Bigg)\sigma_l^4(k)\;.
\end{eqnarray}
Here we assumed that the main contribution to the integrals in (\ref{eq:largek2}) comes  from the integration range $p,q<k$, which 
given the form (\ref{eq:EH_spectrum}) is valid at high $k$ up to logarithmic corrections.
Besides, in the last line we substituted $f(p/q)\sim 1$, which is valid for
$p\ll q$ (or, equivalently, $q\ll p$). One can easily get convinced that this estimate is adequate for 
$f(x)$ given by (\ref{eq:fx}). 

It is straightforward to extend the rearrangement to $n$ loops.
The $n$-loop contribution given in Eq.~(\ref{eq:n-loop}) can be rewritten by integrating over $k_1$.
Due to the symmetry in
the momenta $k_1,\dots k_m$, one may restrict the range of integration to the case where $|k_1|$ is larger than any of the $|k_i|$ for $1<i\leq m$, and multiply by a factor $m$, 
\bea\label{eq:n-loop-theta}
P_{n-loop}(k,\eta) &=& \sum_{m=1}^{n+1} m \int d^3k_2 \, \dots d^3k_{n+1} \,
A^{(n)}_m(\vec k_1, \dots , \vec k_{n+1}) \, \nn \\
&& \times \Theta(|k_1|-|k_2|)\times\dots\times\Theta(|k_1|-|k_m|) \nn \\
&& \times P^L(k_1,\eta) \dots P^L(k_{n+1},\eta)\big|_{\vec k_1=\vec k-\vec k_2\dots -\vec k_m}  \,.
\eea
The integration kernels in the notation of Eq.~(\ref{eq:n-loop}) are given by
\begin{align}
\label{eq:Anm}
 A^{(n)}_m(\vec k_1,\dots,\vec k_{n+1}) &= \sum_{n_L=0}^{n-m+1} \frac{(2n_L+m)!(2n_R+m)!}{2^{n_L+n_R} m! n_L! n_R!} \nn \\
 & \hspace{-.4cm}\times F^{s}_{2n_L+m}(\vec k_1,\dots,\vec k_m,\vec p_1,-\vec p_1,\dots,\vec p_{n_L},-\vec p_{n_L}) \nn\\
 & \hspace{-.4cm}\times F^{s}_{2n_R+m}(\vec k_1,\dots,\vec k_m,\vec q_1,-\vec q_1,\dots,\vec q_{n_R},-\vec q_{n_R}) \,,
\end{align}
where $n_R\equiv n+1-m-n_L$, $\vec p_i\equiv \vec k_{m+i}$, $\vec q_i\equiv \vec k_{m+n_L+i}$. In the usual $P_{IJ}$ notation, a given term in
the sum in (\ref{eq:Anm})  contributes to $P_{2n_L+m,2n_R+m}$. Diagrammatically, a given summand corresponds to
a diagram with two `blobs'  that are connected by $m$ lines, and have $n_L$ and $n_R$ lines connected to themselves.
By adding and subtracting the terms obtained from a Taylor expansion of $P^L(k_1,\eta_0)$ around $k$ up to order $2n$,
one obtains the rearrangement analogous to the one- and two-loop case discussed above.
The coefficients of the terms containing the Taylor-expanded power spectrum  in (\ref{eq:P_asympt}) after the rearrangement are
\be\label{eq:BnlAveraged}
B^{(n)}_l(k,k_1,\dots,k_n) \equiv \frac{1}{(4\pi)^n}\int d\Omega_{k_1}\cdots\int d\Omega_{k_n} B^{(n)}_l(\vec k,\vec k_1,\dots,\vec k_n) \,,
\ee
where
\bea
\label{eq:Bnl}
  B^{(n)}_l(\vec k,\vec k_1,\dots,\vec k_n) &=& A^{(n)}_1(\vec k,\vec k_1,\dots,\vec k_{n})\delta_{l0}  \nn\\
&& + \sum_{m=2}^{n+1} m A^{(n)}_m(\vec k-\vec k_1\dots-\vec k_{m-1},\vec k_1,\dots,\vec k_{n})\nn\\
&& \times b_l(\vec k,\vec k_1+\dots+\vec k_{m-1})\Big|_{symm}\,.
\eea
Here, the right-hand side is to be fully symmetrized w.r.t.~permuting the $\vec k_i$ and w.r.t.~inverting the sign $\vec k_i\to -\vec k_i$ of each momentum. We note that in our numerical results (up to four loop order) we find that the cancellation of terms growing with $k$ can even be observed at the level of the non-averaged expression.

\section{The power spectrum in the soft regime\label{sec:app:soft}}

In the limit $k\to 0$, the dominant contribution to the $n$-loop correction to
the power spectrum is
given by \cite{Fry:1993bj,Valageas:2001td}
\begin{eqnarray}
 P_{n-loop}(k) &\to&2\, (2n+1)!!\, P^L(k) \int_{q_1}\cdots\int_{q_n} F_{2n+1}^s(\vec
k,\vec q_1,-\vec q_1,\dots,\vec q_n,-\vec q_n) \nonumber\\
&& {} \times P^L(q_1)\cdots P^L(q_n) \;.
\label{eq:LowkP}
\end{eqnarray}
Due to the well-known property $F_{2n+1}^s\propto k^2$ of the PT kernels
\cite{Goroff:1986ep}, these corrections scale
as $k^2P^L(k)$. All other loop corrections would lead to terms scaling as $k^4$
or $k^4P^L(k)$, respectively,
that are parametrically smaller for $k\to 0$ for a power spectrum $P^L(k)\propto
k^{n_s}$ with $n_s\lesssim 1$. The function appearing in the two-loop correction
(\ref{smallk}) is given by
\begin{eqnarray}
 g(x) &=& \frac{1}{179056
x^6}\Bigg((x^2+1)\left(128258x^4-5760(x^8+1)-13605(x^6+x^2)\right)\nonumber\\
 && {} -
\frac{15}{4x}(x^2-1)^4\left(384(x^4+1)+2699x^2\right)\ln\left(\frac{x-1}{x+1}
\right)^2 \Bigg)\;.
\end{eqnarray}

\section{Convergence and fictitious power spectra\label{sec:app:conv}}

\begin{figure}[t!]
\begin{center}
\includegraphics[width=0.9\textwidth, clip ]{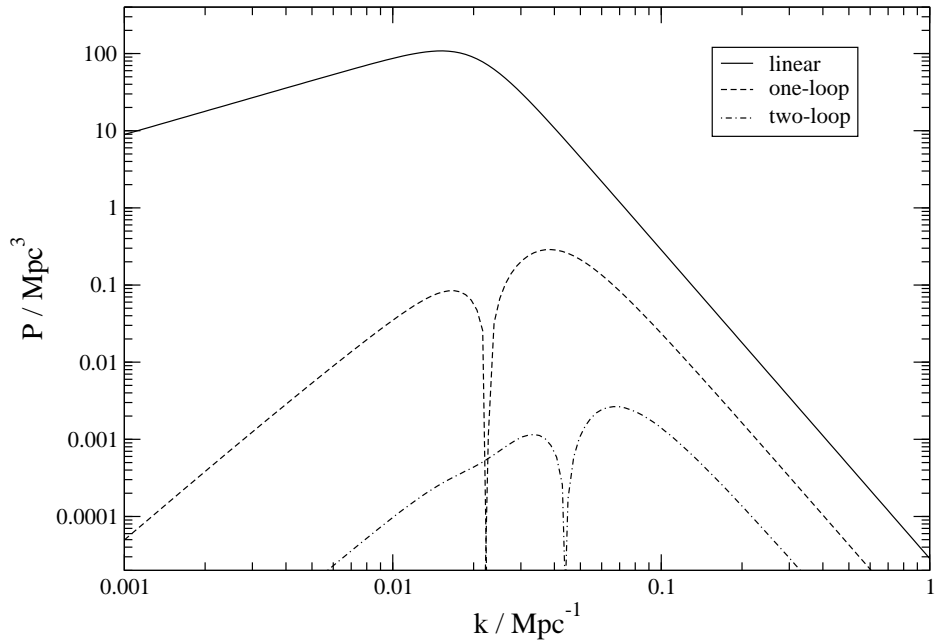}
\end{center}
\caption{
\label{fig:spec_conv}
\small Example for a power spectrum where $\sigma_l^2$ is not sensitive to the high momentum region. The different lines are the linear, one-loop and two-loop contributions. The corresponding parameters are $k_0^2 \sigma_d^2 \simeq 0.015$ and $\sigma_l^2 \simeq 0.02$. }
\end{figure}

In this section, we present a two-loop result for a fictitious power spectrum with better convergence behavior
than (\ref{eq:EH_spectrum}).  Let us consider an initial power spectrum of the form
\be
P_0( k ) \propto \frac{k\, k_0}{k_0^5 + k^5} \, .
\ee
This spectrum is chosen such that the expansion parameter of standard perturbation theory $\sigma_l^2$ is not sensitive to the high momentum part of the spectrum. Hence $\sigma_l^2$ is small as long as $k_0^2 \sigma_d^2$ is much smaller than unity. Fig.~\ref{fig:spec_conv} shows a power spectrum with $k_0^2 \sigma_d^2 \simeq 0.015$ and $\sigma_l^2 \simeq 0.02$. Even though $k_0^2 \sigma_d^2$ is as large as in the physical case at $z\sim 0$, the two-loop result is well below the one-loop contribution and standard perturbation theory is expected to converge. Compare with Fig.~\ref{fig:wmap5}.

\end{appendix}


\end{document}